\documentclass[journal]{IEEEtran}
\usepackage{amssymb,amsfonts}
\usepackage[cmex10]{amsmath}

\usepackage{multirow}

\usepackage{graphicx}%
\usepackage{mathrsfs}%
\usepackage{xcolor}%
\usepackage{textcomp}%
\usepackage{manyfoot}%
\usepackage{booktabs}%
\usepackage{algorithm}%
\usepackage{algorithmicx}%
\usepackage{algpseudocode}%
\usepackage{listings}%
\usepackage{lmodern}

\newcommand{\KQ}[1]{\textcolor{red}{#1}}

\usepackage{svg}
\usepackage[switch,pagewise]{lineno}
\usepackage{soul}
\usepackage{placeins} 
\usepackage{hyperref}
\usepackage{array}
\usepackage{float}
\usepackage{url}

\begin{document}
%
\title{Geometry-Decoupled Deep Unfolding for Gridless Super-Resolution TomoSAR Under Nonuniform Baselines}
%
%
%

\author{Kun~Qian, Zhuge Xia, Qian Ma, Qi Zhang, and Xiufeng He, \IEEEmembership{Senior Member, IEEE}
\thanks{This work was supported by the Basic Research Program of Jiangsu under Grant BK20251497, the China Postdoctoral Foundation under Grants 2025M784522 and 2025M780213, and the National Natural Science Foundation of China under Grant 62501636. (\textit{Corresponding author}: Zhuge Xia.)}
\thanks{Kun Qian is with Wuhan Electronic Information Institute, Wuhan 430019, China (e-mail: kun\_qian\_academic@163.com).}%
\thanks{Zhuge Xia and Xiufeng He are with the School of Earth Sciences and Engineering, Hohai University, Nanjing 211100, China (e-mail: zhugexia@hhu.edu.cn; xfhe@hhu.edu.cn).}  
\thanks{Qian Ma is with the School of Medical Information Engineering, Guangzhou University of Chinese Medicine, Guangzhou, China (e-mail: maqian@gzucm.edu.cn).}
\thanks{Qi Zhang is with the Chair of Data Science in Earth Observation, Technical University of Munich, 80333 Munich, Germany (e-mail: rachelqi.zhang@tum.de).}
}

\maketitle


\begin{abstract}
\KQ{(This work has been submitted to the IEEE for possible publication. Copyright may be transferred without notice, after which this version may no longer be accessible.)}Super-resolution SAR tomography (TomoSAR) is performed on a discretized elevation grid, leading to off-grid bias and spectral leakage. Classical Toeplitz-Vandermonde gridless formulations avoid elevation discretization but rely on uniform sampling, whereas covariance- or subspace-based estimation is difficult in single-look repeat-pass TomoSAR. We propose DUSG-Tomo-Net, a geometry-decoupled deep unfolding framework for gridless inversion under nonuniform baselines. In DUSG-Tomo-Net, pairwise products of a single observation vector are related to a latent Toeplitz-compatible virtual-lag sequence through an analytical geometry operator and modeled as noisy covariance surrogates containing multi-scatterer cross terms, measurement noise, and lag-interpolation errors. Each unfolded layer combines learned lag-domain regularization, closed-form geometry-dependent data consistency, and finite-step Dykstra projection toward the set of Hermitian Toeplitz positive semidefinite matrices. Root-MUSIC then retrieves scatterer elevations in the continuous domain without an elevation dictionary. Because the trainable modules operate only on the common virtual-lag representation, the learned parameterization can be reused across representationally compatible acquisition geometries by recomputing the geometry operator. At 6~dB, simulations under nonuniform baselines yield a single-scatterer RMSE of 0.703~m and an effective detection rate of approximately 0.90 for two scatterers at the Rayleigh limit, while demonstrating reliable sub-Rayleigh separation. A model trained with 20 acquisitions remains applicable to perturbed configurations containing 8-28 acquisitions without retraining. Experiments on a 16-image CH-1 stack show that DUSG-Tomo-Net produces coherent urban elevation maps and more continuous double-scatterer structures than the gridless baseline method TADCG. The same CH-1-trained network preserves the dominant elevation structure when applied to 12- and 8-image subsets without retraining.
\end{abstract}


%
\IEEEpeerreviewmaketitle

\newcommand{\bA}{\mathbf{A}}
\newcommand{\bT}{\mathbf{T}}
\newcommand{\bc}{\mathbf{c}}
\newcommand{\bg}{\mathbf{g}}
\newcommand{\bn}{\mathbf{n}}
\newcommand{\by}{\mathbf{y}}
\newcommand{\bR}{\mathbf{R}}
\newcommand{\bx}{\mathbf{x}}
\newcommand{\ba}{\mathbf{a}}
\newcommand{\bZ}{\mathbf{Z}}
\newcommand{\bu}{\mathbf{u}}
\newcommand{\bv}{\mathbf{v}}
\newcommand{\bU}{\mathbf{U}}
\newcommand{\bPsi}{\boldsymbol{\Psi}}
\newcommand{\bvarepsilon}{\boldsymbol{\varepsilon}}
\newcommand{\bgamma}{\boldsymbol{\gamma}}
\newcommand{\bxi}{\boldsymbol{\xi}}
\newcommand{\bs}{\mathbf{s}}
\newcommand{\bD}{\mathbf{D}}
\newcommand{\bI}{\mathbf{I}}
\newcommand{\cA}{\mathcal{A}}
\newcommand{\bX}{\mathbf{X}}
\newcommand{\bP}{\mathbf{P}}

\section{Introduction}
\IEEEPARstart{S}{ynthetic} aperture radar (SAR) provides all-weather, day-and-night imaging capability for large-scale Earth observation. However, conventional SAR projects the three-dimensional (3-D) scattering structure of a scene onto a two-dimensional azimuth-range image, causing scatterers at different elevations to be superimposed within the same resolution cell and giving rise to the layover effect. SAR tomography (TomoSAR) resolves this elevation ambiguity by exploiting multibaseline acquisitions to synthesize an aperture along the elevation dimension \cite{Reigber2000, Lombardini2005, Fornaro2014}. This capability has enabled elevation-resolved imaging for urban mapping \cite{Zhu2010superres, Yang2023distributedtomo}, deformation monitoring \cite{Fornaro2009}, forest profiling \cite{Cloude2006, Lu2022phase}, and cryosphere observation \cite{wu2011ice}. It is particularly important in dense urban environments, where echoes from building facades, rooftops, and ground surfaces are frequently superimposed within the same azimuth-range resolution cell and must be separated along the elevation dimension for reliable 3-D reconstruction.

In repeat-pass spaceborne TomoSAR, orbital constraints generally limit both the effective cross-track aperture and the number of coherent acquisitions. The finite elevation aperture therefore yields a Rayleigh elevation resolution that is often on the order of tens of meters \cite{Zhu2010superres, Zhu2011tomo}, making conventional Fourier-based estimators inadequate for resolving closely spaced scatterers in dense urban scenes. This limitation motivates super-resolution TomoSAR inversion. In urban areas, the elevation reflectivity profile within an individual azimuth-range resolution cell is typically dominated by only a few strong scattering contributions \cite{Zhu2010superres, Zhu2011tomo}, providing a natural sparsity prior for super-resolution processing.

Super-resolution TomoSAR has been studied from complementary detection- and reconstruction-oriented perspectives. Detection-oriented approaches employ statistical hypothesis testing, particularly generalized likelihood ratio test (GLRT)-based schemes, to determine scatterer presence and model order \cite{DeMaio2009glrt, Pauciullo2012double, Budillon2016supglrt, Haddad2022beyond, Haddad2024gridlessglrt}. In contrast, reconstruction-oriented approaches recover the elevation reflectivity profile or its spectral support, with compressive sensing (CS) being one of the most widely adopted formulations for exploiting elevation sparsity. Representative CS-TomoSAR methods, including greedy pursuit algorithms and $\ell_1$-regularized sparse reconstruction, have demonstrated substantial improvements in elevation resolution and localization accuracy \cite{Zhu2010superres, Budillon2011, REBMEISTER2021220}. However, most CS-based TomoSAR methods reconstruct the reflectivity profile on a predefined elevation grid. A mismatch between the continuous scatterer locations and the discretized dictionary can introduce spectral leakage and localization bias, whereas grid refinement reduces the discretization scale only at the expense of increased dictionary coherence and computational cost.

Gridless methods avoid elevation discretization by estimating scatterer locations directly in the continuous domain. Atomic norm minimization (ANM) is a representative framework that replaces the finite-dimensional sparse prior with a continuously parameterized atomic norm \cite{Tang2013anm, Bhaskar2013anm, Wang2022anm}. However, repeat-pass spaceborne TomoSAR poses two structural challenges for classical gridless formulations. First, the Toeplitz-Vandermonde representation underlying classical ANM relies on uniformly sampled observations with a uniform linear array (ULA) structure, whereas the physical perpendicular baselines of repeat-pass acquisitions are generally nonuniform. Consequently, the raw TomoSAR measurement vector does not directly exhibit the uniform Vandermonde structure required by the classical formulation. A second challenge arises in the single-look regime, where reliable covariance or subspace estimation is difficult because only one complex observation vector is available per resolution cell in high-resolution urban TomoSAR.

Existing strategies for gridless TomoSAR under nonuniform sampling follow different routes. One route maps irregular measurements or their lag-domain information to a uniform virtual-lag representation before applying structured gridless reconstruction. Such approaches recover a Toeplitz-compatible representation, but introduce an additional geometry-dependent approximation stage and may rely on sampling redundancy or specific acquisition conditions \cite{Gao2024}. Another route operates directly on nonuniform measurements by optimizing continuously parameterized scatterer models, including TADCG \cite{Shao2024TADCG}, GDLS \cite{Shi2023gdls}, and ANM-based formulations developed for nonuniform sampling \cite{Liu2024nonuniformanm}. These methods remove the uniform-aperture restriction, but their inference remains iterative and method-specific, with computational behavior and algorithmic parameters tied to the underlying optimization procedure. More fundamentally, under nonuniform baselines, both the measurement operator and the associated inversion procedure vary with the number and distribution of acquisitions, making acquisition geometry an explicit component of the reconstruction problem.

Deep algorithm unrolling \cite{Gregor2010lista, Monga2021unrolling} provides a model-driven approach to accelerating iterative inverse solvers by mapping a finite number of optimization iterations into trainable network layers while retaining explicit data-consistency and model-based operations. This paradigm has motivated a series of learned TomoSAR reconstruction methods \cite{Qian2022gammanet, Qian2022bpdnrnn, Wang2023atasi, Wang2023madanet}, which have demonstrated competitive super-resolution performance with substantially reduced inference cost. However, existing unfolded TomoSAR methods are predominantly formulated on discretized elevation dictionaries and therefore retain the off-grid limitation of grid-based reconstruction.

Beyond this discretization issue, a central obstacle to transferring learned TomoSAR inversion across acquisition stacks is its coupling to the acquisition geometry. In many unfolded architectures, the sensing matrix or a geometry-dependent learned surrogate is embedded directly in the layerwise transformations. Consequently, changing the number or distribution of baselines modifies the measurement operator and can invalidate the geometry-specific transformations learned during training, often necessitating retraining or model adaptation. Analytical weight determination, as adopted in HyperLISTA-ABT \cite{Qian2024hyperlista}, substantially reduces the number of trainable parameters and the associated training burden, but the reconstruction remains defined on a geometry-dependent discretized elevation dictionary. More recently, GITomo-Net \cite{Liu2025gitomonet} has addressed acquisition-geometry variation by learning shared representations across different baseline configurations, representing an important step toward transferable learned TomoSAR reconstruction. Nevertheless, its reconstruction remains grid-based and therefore retains its dependence on a predefined elevation grid and the associated off-grid mismatch. Consequently, geometry adaptation and continuous-domain gridless reconstruction have largely been studied as separate problems in learned TomoSAR inversion. A unified unfolded formulation that accommodates varying nonuniform baseline geometries while retaining continuous-domain elevation estimation remains largely unexplored.

To bridge geometry adaptation and continuous-domain gridless reconstruction, we propose DUSG-Tomo-Net, a geometry-decoupled deep unfolding framework for super-resolution TomoSAR under nonuniform baselines. The proposed method maps pairwise products of the single-look observations to a Toeplitz-compatible virtual-lag representation through an analytically constructed geometry operator. DUSG-Tomo-Net then unfolds a structured lag-domain estimator that combines learned lag-domain regularization, geometry-dependent data consistency, and Toeplitz-PSD structural enforcement. The recovered structured representation is finally processed by Root-MUSIC to estimate scatterer elevations directly in the continuous domain. By separating the analytical measurement mapping from the trainable lag-domain regularization, the same learned parameterization can be reused across compatible acquisition geometries. The main contributions of this paper are summarized as follows:
\begin{itemize}
    \item We formulate single-look TomoSAR inversion under nonuniform baselines as structured estimation in a Toeplitz-compatible virtual-lag domain. Per-pixel pairwise products are modeled as noisy lag-domain covariance surrogates, enabling continuous-domain spectral estimation without an elevation dictionary.

    \item We introduce a geometry-decoupled unfolding strategy that separates acquisition-specific measurement mapping from geometry-shared learned regularization. This formulation enables the same trainable parameterization to be reused across representationally compatible baseline configurations by updating the geometry operator rather than retraining the network.

    \item We develop an unfolded estimator that combines learned lag-domain regularization, closed-form geometry-dependent data consistency, and projection-based Toeplitz-PSD enforcement. The residual terms arising from the single-look pairwise-product approximation are explicitly decomposed in the resulting observation model.
\end{itemize}

The remainder of this paper is organized as follows. Section~II presents the TomoSAR signal model and reviews grid-based and gridless inversion. Section~III develops the virtual-lag formulation and the proposed geometry-decoupled unfolded framework. Section~IV presents simulation and real-data experiments. Section~V discusses practical applicability and limitations. Section~VI concludes the paper.

\section{Imaging Model and Background}
\subsection{TomoSAR Imaging Model}

\begin{figure}[h]
  \centering
  \includegraphics[width=0.95\linewidth]{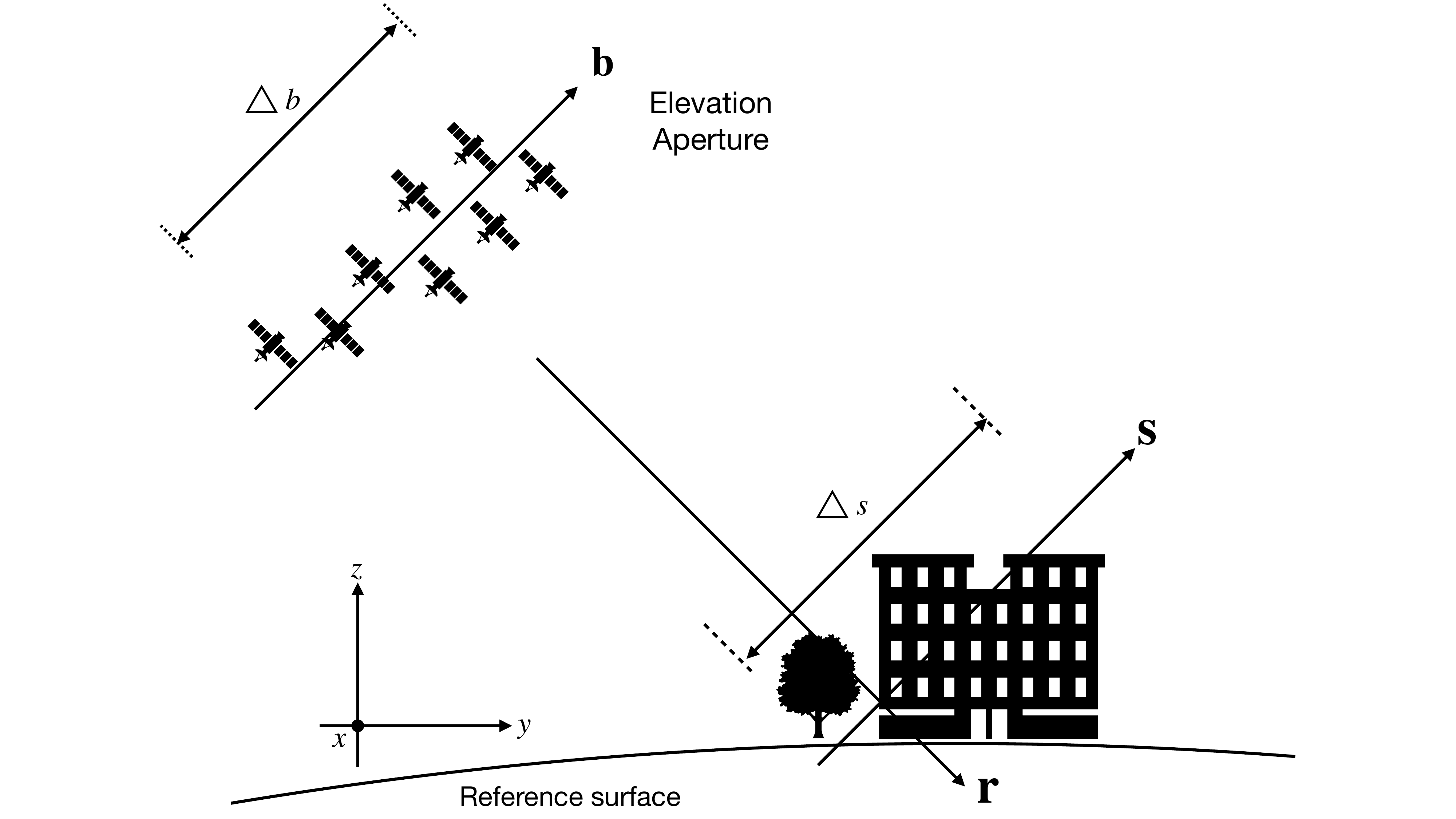}
  \caption{TomoSAR imaging geometry. The elevation synthetic aperture is formed by coherent SAR acquisitions collected from slightly different viewing angles. The flight direction is perpendicular to the plane of the figure.}
  \label{fig:tomo_geometry}
\end{figure}

TomoSAR reconstructs the elevation reflectivity profile by exploiting the baseline-dependent phase diversity of multibaseline SAR acquisitions (see Fig.~\ref{fig:tomo_geometry}) and can be interpreted as a spectral estimation problem along the elevation dimension \cite{Reigber2000, Lombardini2005, Zhu2011tomo}. For a fixed azimuth-range resolution cell observed by $N$ coherent acquisitions with perpendicular baselines $\{b_n\}_{n=1}^{N}$ relative to a reference track, the measurement at the $n$th acquisition is modeled as
\begin{equation}
  g_n = \int_{\mathcal{S}} \gamma(s)
  \exp\!\Bigl(\frac{j4\pi b_n s}{\lambda r}\Bigr)\, ds + \varepsilon_n,
  \label{eq:tomo_cont}
\end{equation}
where $\gamma(s)$ denotes the continuous elevation reflectivity profile, $\mathcal{S}$ denotes the elevation support of the illuminated scene, $s_{\max}\triangleq\max_{s\in\mathcal{S}}|s|$ is the maximum elevation magnitude within this support, $\lambda$ is the radar wavelength, $r$ is the slant range, and $\varepsilon_n$ denotes additive complex Gaussian noise. Equation~\eqref{eq:tomo_cont} shows that the TomoSAR observation is a Fourier-type measurement whose sampling locations along the elevation-frequency dimension are determined by the physical perpendicular baselines. For repeat-pass spaceborne acquisitions, these baseline samples are generally nonuniform.

For grid-based reconstruction, the continuous elevation support can be discretized using $L$ grid points $\{s_l\}_{l=1}^{L}$ with spacing $\delta s$, yielding the approximation
\begin{equation}
  g_n \approx \delta s \sum_{l=1}^{L} \gamma(s_l)
  \exp\!\Bigl(\frac{j4\pi b_n s_l}{\lambda r}\Bigr) + \varepsilon_n.
  \label{eq:tomo_disc}
\end{equation}
Defining $\gamma_l \triangleq \delta s\,\gamma(s_l)$ and stacking the $N$ measurements gives
\begin{equation}
  \bg = \bR \bgamma + \bvarepsilon,
  \label{eq:tomo_vec}
\end{equation}
where $\bg \in \mathbb{C}^{N}$ is the observation vector, $\bgamma = [\gamma_1,\ldots,\gamma_L]^{\top} \in \mathbb{C}^{L}$ is the discrete elevation reflectivity vector on the predefined grid, and $\bvarepsilon \sim \mathcal{CN}(\mathbf{0},\sigma^2\bI)$. The steering matrix $\bR \in \mathbb{C}^{N\times L}$ is determined jointly by the physical baseline configuration and the prescribed elevation grid, with entries
\begin{equation}
  R_{nl} = \exp\!\Bigl(\frac{j4\pi b_n s_l}{\lambda r}\Bigr).
  \label{eq:steering_entry}
\end{equation}
The discretized model in Eq.~\eqref{eq:tomo_vec} provides the standard representation used by grid-based TomoSAR reconstruction methods. In the high-resolution urban setting considered in this work, spatial multilooking is avoided and only one complex observation vector $\bg$ is available for each azimuth-range resolution cell. Moreover, the elevation response within an individual cell is typically dominated by only a few strong scattering contributions. These characteristics motivate the sparse grid-based formulation reviewed next, while the continuous model in Eq.~\eqref{eq:tomo_cont} provides the basis for the subsequent gridless formulation.

\subsection{Grid-Based TomoSAR Inversion via Sparse Reconstruction}

Based on the discretized imaging model in Eq.~\eqref{eq:tomo_vec}, compressive sensing (CS)-based TomoSAR (CS-TomoSAR) represents the elevation reflectivity profile on the predefined grid $\{s_l\}_{l=1}^{L}$ as a sparse coefficient vector $\bgamma\in\mathbb{C}^{L}$ and formulates per-pixel inversion as a sparse reconstruction problem \cite{Zhu2010superres, Zhu2011tomo, Budillon2011}. A standard formulation is the $\ell_1$-regularized least-squares problem
\begin{equation}
  \hat{\bgamma}
  = \arg\min_{\bgamma}
  \frac{1}{2}\|\bg - \bR\bgamma\|_2^2
  + \mu \|\bgamma\|_1,
  \label{eq:l1_tomosar}
\end{equation}
where $\mu>0$ balances data fidelity and sparsity. Equation~\eqref{eq:l1_tomosar} is closely related to the basis pursuit denoising (BPDN) and LASSO formulations \cite{Chen1998bp, lasso} and can be solved using proximal-gradient, alternating-direction, or interior-point algorithms \cite{Daubechies2004ista, Beck2009fista, wright1997primal}. Representative TomoSAR implementations include SL1MMER \cite{Zhu2011tomo} and related sparse reconstruction frameworks \cite{Zhu2010superres, Budillon2011}.

Grid-based sparse reconstruction can substantially improve elevation resolution over conventional Fourier-type spectral estimators. However, its accuracy depends on how well the predefined elevation dictionary represents the true continuous scatterer locations. Since physical scatterer elevations generally do not coincide exactly with the grid points, off-grid scatterers must be represented by neighboring dictionary atoms, giving rise to basis mismatch that can cause energy leakage, localization bias, and spurious responses \cite{Chi2011}. For a fixed elevation grid, this discretization-induced bias does not necessarily vanish with increasing SNR. Refining the grid reduces the quantization scale but simultaneously increases the coherence between neighboring dictionary atoms, deteriorates numerical conditioning, and raises the computational cost. Grid refinement therefore alleviates, but does not fundamentally eliminate, the model mismatch introduced by elevation discretization. This limitation motivates continuous-domain gridless TomoSAR formulations.

\subsection{Gridless TomoSAR and Challenges Under Nonuniform Baselines}

Gridless methods avoid elevation discretization by parameterizing scatterer locations directly in the continuous domain \cite{Tang2013anm, Bhaskar2013anm, Wang2022anm, Yang2015gridless, Liu2022urbananm}. Atomic norm minimization (ANM) provides a representative convex framework for continuous spectral estimation.

Consider first a uniformly sampled elevation aperture
\begin{equation}
  b_n=b_0+(n-1)\Delta_{\mathrm{u}},
  \qquad n=1,\ldots,N,
  \label{eq:ula_baseline}
\end{equation}
where $\Delta_{\mathrm{u}}$ denotes the uniform baseline spacing. Assuming $K$ point-like scatterers, the phase term associated with the reference baseline $b_0$ can be absorbed into the complex scatterer coefficients by defining $\tilde{\gamma}_k \triangleq \gamma_k \exp\!\left( \frac{j4\pi b_0 s_k}{\lambda r} \right)$.
The noiseless observation vector can then be expressed as
\begin{equation}
  \bx
  =
  \sum_{k=1}^{K}
  \tilde{\gamma}_k\ba(f_k),
  \label{eq:anm_signal}
\end{equation}
where $f_k=\frac{2\Delta_{\mathrm{u}}\,s_k}{\lambda r}$ is the normalized elevation frequency of the $k$th scatterer, represented on the signed principal interval $[-1/2,1/2)$. The corresponding Vandermonde atom is
\begin{equation}
  \ba(f)
  =
  \left[
    1,
    e^{j2\pi f},
    \ldots,
    e^{j2\pi(N-1)f}
  \right]^{\top}
  \in\mathbb{C}^{N}.
  \label{eq:atom_vec}
\end{equation}
Defining the continuously parameterized atomic set
\begin{equation}
  \mathcal{A}
  =
  \left\{
    \ba(f): f\in[-1/2,1/2)
  \right\},
  \label{eq:atomic_set}
\end{equation}
the associated atomic norm is
\begin{equation}
  \|\bx\|_{\mathcal{A}}
  =
  \inf
  \left\{
    \sum_{k}|\alpha_k|:
    \bx=\sum_{k}\alpha_k\ba(f_k),
    \ f_k\in[-1/2,1/2)
  \right\}.
  \label{eq:atomic_norm}
\end{equation}
The atomic norm therefore promotes a sparse spectral representation while retaining continuous-valued scatterer locations.

Given the noisy observation $\bg=\bx+\bvarepsilon$, atomic norm denoising can be formulated as
\begin{equation}
  \hat{\bx}
  =
  \arg\min_{\bx}
  \frac{1}{2}\|\bg-\bx\|_2^2
  +
  \tau\|\bx\|_{\mathcal{A}},
  \label{eq:anm_basic}
\end{equation}
where $\tau>0$ controls the regularization strength. Using the Carath\'eodory-Toeplitz representation \cite{Caratheodory-Toeplitz}, Eq.~\eqref{eq:anm_basic} admits the semidefinite formulation
\begin{equation}
\begin{aligned}
  \min_{\bx,\bc,t}\quad
  &
  \frac{1}{2}\|\bg-\bx\|_2^2
  +
  \tau
  \left[
    \frac{1}{2N}
    \mathrm{Tr}\!\bigl(\bT(\bc)\bigr)
    +
    \frac{t}{2}
  \right]
  \\
  \text{s.t.}\quad
  &
  \begin{bmatrix}
    \bT(\bc) & \bx\\
    \bx^{H} & t
  \end{bmatrix}
  \succeq 0,
\end{aligned}
  \label{eq:anm_sdp}
\end{equation}
where $\bc\in\mathbb{C}^{N}$ parameterizes the first column of the Hermitian Toeplitz matrix $\bT(\bc)\in\mathbb{C}^{N\times N}$, and $t\in\mathbb{R}_{+}$ is an auxiliary variable \cite{Bhaskar2013anm, Tang2013anm, Wang2022anm}. A rank-$K$ Hermitian Toeplitz positive semidefinite matrix with $K<N$ admits a Vandermonde decomposition whose atoms encode the continuous spectral locations.

The above Toeplitz-Vandermonde formulation critically relies on uniform baseline spacing. However, in repeat-pass spaceborne TomoSAR, the perpendicular baselines are generally nonuniform, and the physical observation vector therefore does not directly exhibit the uniform Vandermonde structure required by classical ANM. To recover a Toeplitz-compatible representation, the irregular measurements must either be reformulated on a uniform virtual-lag domain or handled by a gridless formulation designed directly for nonuniform sampling. The former route, which is relevant to this work, exploits baseline-difference information to construct a structured second-order representation from pairwise products of the measurements. Nevertheless, this reformulation introduces a second difficulty in the high-resolution single-look setting. In urban TomoSAR, spatial multilooking is commonly avoided to preserve the native azimuth-range resolution, so that only one complex observation vector is available for each resolution cell. Consequently, pairwise products formed from a single realization cannot be regarded as stable ensemble covariance samples but only as realization-dependent covariance surrogates. The desired Toeplitz-structured representation therefore cannot be obtained through direct covariance construction and must instead be estimated under appropriate structural constraints. A third difficulty concerns the inference procedure itself. Classical ANM and many gridless methods for nonuniform measurements rely on convergence-driven iterative optimization, such as semidefinite programming or method-specific continuous-parameter updates. Their computational cost and inference behavior consequently depend on the adopted solver, convergence rate, and stopping criteria.

Taken together, gridless TomoSAR under nonuniform baselines involves two structural challenges: restoring a Toeplitz-compatible representation from irregular baseline sampling and estimating that representation reliably from single-look covariance surrogates. In addition, conventional gridless reconstruction commonly relies on convergence-driven iterative optimization, introducing a practical limitation in computational efficiency. These issues motivate the virtual-lag structured estimation and deep unfolding framework developed in Section~III.

\section{Proposed DUSG-Tomo-Net}
This section develops DUSG-Tomo-Net from its structured formulation to the unfolded reconstruction architecture. We first reformulate the nonuniform-baseline observations in a Toeplitz-compatible virtual-lag domain and derive a model-based proximal-projection estimator for the latent lag sequence. We then unfold the resulting iterative scheme into the proposed geometry-decoupled network and present the continuous-domain spectral readout and training objective.

\subsection{Virtual-Lag Formulation Under Nonuniform Baselines}

We first reformulate the nonuniform-baseline TomoSAR observations in a common virtual-lag domain. Let $\mathcal{G}=\{b_n\}_{n=1}^{N}$ denote the acquisition geometry specified by the physical perpendicular baselines. For a resolution cell containing $K$ point-like scatterers, we represent the continuous reflectivity profile as
$\gamma(s)=\sum_{k=1}^{K}\gamma_k\delta(s-s_k)$, where $\gamma_k$ denotes the complex scattering coefficient of the $k$th scatterer. The noiseless component of the observation in Eq.~\eqref{eq:tomo_cont} can then be written as
\begin{equation}
  x_n
  =
  \sum_{k=1}^{K}
  \gamma_k
  \exp\!\left(
    \frac{j4\pi b_n s_k}{\lambda r}
  \right),
  \qquad n=1,\ldots,N,
  \label{eq:point_scatterer_obs}
\end{equation}
such that $g_n=x_n+\varepsilon_n$. For a pair of acquisitions $(m,n)$, the product of the noiseless measurements admits the exact decomposition
\begin{equation}
\begin{aligned}
  x_m x_n^{*}
  &=
  \sum_{k=1}^{K}
  |\gamma_k|^2
  \exp\!\left[
    \frac{j4\pi (b_m-b_n)s_k}{\lambda r}
  \right]
  \\
  &\quad+
  \sum_{\substack{k,l=1\\k\neq l}}^{K}
  \gamma_k\gamma_l^{*}
  \exp\!\left[
    \frac{j4\pi (b_m s_k-b_n s_l)}{\lambda r}
  \right].
  \label{eq:pairwise_decomposition}
\end{aligned}
\end{equation}
The first term depends only on the baseline difference
$\tau_{mn}=b_m-b_n$, whereas the second term depends jointly on the two absolute baseline positions. We therefore define the desired lag-dependent component as
\begin{equation}
  r(\tau)
  \triangleq
  \sum_{k=1}^{K}
  |\gamma_k|^2
  \exp\!\left(
    \frac{j4\pi \tau s_k}{\lambda r}
  \right).
  \label{eq:ideal_lag_response}
\end{equation}
The second term in Eq.~\eqref{eq:pairwise_decomposition}, denoted by $\varepsilon_{\mathrm{cross},mn}$ hereafter, represents the realization-dependent multi-scatterer cross terms.

To obtain a Toeplitz-compatible representation, we introduce a uniform virtual-lag grid $\mathcal{T}_{v}=\{q\Delta_b\}_{q=0}^{N_v-1}$, where $\Delta_b$ denotes the virtual-lag spacing and $N_v$ is the number of virtual-lag samples. Sampling the desired lag response in Eq.~\eqref{eq:ideal_lag_response} on this grid defines $c_q \triangleq r(q\Delta_b)$ for $q=0,\ldots,N_v-1$ and yields
\begin{equation}
  \bc
  =
  [c_0,c_1,\ldots,c_{N_v-1}]^{\top}
  \in\mathbb{C}^{N_v}.
  \label{eq:virtual_lag_coeff}
\end{equation}
We define the virtual-domain steering vector
\begin{equation}
  \ba_v(s)
  =
  \left[
  1,\,
  e^{j\frac{4\pi\Delta_b s}{\lambda r}},\,
  \ldots,\,
  e^{j\frac{4\pi(N_v-1)\Delta_b s}{\lambda r}}
  \right]^{\top}.
  \label{eq:virtual_steering}
\end{equation}
Because $\ba_v(s)$ is periodic in elevation with period $\lambda r/(2\Delta_b)$, the prescribed elevation support must lie within a single principal phase interval for unambiguous spectral localization. Accordingly, we require $\mathcal{S} \subset \left[-\frac{\lambda r}{4\Delta_b}, \frac{\lambda r}{4\Delta_b}\right)$. For a symmetric closed support $|s|\leq s_{\max}$, a sufficient condition is $\Delta_b<\lambda r/(4s_{\max})$.

Let $\bT(\bc)\in\mathbb{C}^{N_v\times N_v}$ denote the Hermitian Toeplitz lifting whose nonnegative-lag sequence is parameterized by $\bc$. For the ideal lag sequence defined above, this matrix admits the Vandermonde decomposition
\begin{equation}
  \bT(\bc)
  =
  \sum_{k=1}^{K}
  |\gamma_k|^2
  \ba_v(s_k)\ba_v^{H}(s_k)
  \succeq 0.
  \label{eq:virtual_toeplitz_decomposition}
\end{equation}
Hence, $\bT(\bc)$ is Hermitian Toeplitz and positive semidefinite. If the $K$ scatterer elevations are distinct and $K<N_v$, then $\bT(\bc)$ has rank $K$, and the Carath\'eodory-Toeplitz uniqueness result recalled in Section~II-C ensures that its Vandermonde decomposition is unique up to permutation. Together with the principal-branch condition above, this result implies that the elevations $\{s_k\}_{k=1}^{K}$ are uniquely determined by the ideal structured matrix $\bT(\bc)$. This identifiability provides the basis for the Root-MUSIC readout developed in Section~III-D.

We next relate the nonuniform physical observations to the latent virtual-lag sequence. Define the set of admissible ordered baseline pairs as
\begin{equation}
  \mathcal{P}_{\mathcal{G}}
  =
  \left\{
    (m,n):
    0<b_m-b_n\leq(N_v-1)\Delta_b
  \right\},
  \label{eq:admissible_pairs}
\end{equation}
and let $P=|\mathcal{P}_{\mathcal{G}}|$. For the $p$th pair $(m_p,n_p)\in\mathcal{P}_{\mathcal{G}}$, define
\begin{equation}
  \tau_p
  =
  b_{m_p}-b_{n_p},
  \qquad
  y_p
  =
  g_{m_p}g_{n_p}^{*}.
  \label{eq:pairwise_observation}
\end{equation}
Only positive lags are retained because the corresponding negative-lag coefficients are determined by Hermitian symmetry. Although no zero-lag pair with $\tau_p=0$ is directly observed, $c_0$ can be constrained indirectly through interpolation whenever physical lags fall within the first virtual-lag interval $0<\tau_p<\Delta_b$ and is jointly estimated with the remaining lag coefficients.

For $0<\tau_p\leq(N_v-1)\Delta_b$, we define $d_p = \min\!\left( \left\lfloor\frac{\tau_p}{\Delta_b}\right\rfloor, N_v-2 \right)$ and $\eta_p = \frac{\tau_p}{\Delta_b}-d_p$, where $\eta_p\in[0,1]$ and $\eta_p=1$ occurs only at the upper endpoint $\tau_p=(N_v-1)\Delta_b$. The desired lag response is then approximated by linear interpolation between the two neighboring virtual-lag samples,
\begin{equation}
  r(\tau_p)
  \approx
  (1-\eta_p)c_{d_p}
  +
  \eta_p c_{d_p+1}.
  \label{eq:virtual_lag_interp}
\end{equation}
The interpolation error in Eq.~\eqref{eq:virtual_lag_interp} is defined as $\varepsilon_{\mathrm{interp},p} \triangleq r(\tau_p) - \left[(1-\eta_p)c_{d_p} + \eta_p c_{d_p+1}\right]$.

The interpolation weights for all admissible pairs define the geometry operator $\bPsi_{\mathcal G}\in\mathbb{R}^{P\times N_v}$, whose entries are given by
\begin{equation}
  [\bPsi_{\mathcal G}]_{p,q}
  =
  \begin{cases}
    1-\eta_p,
      & q=d_p,\\
    \eta_p,
      & q=d_p+1,\\
    0,
      & \text{otherwise},
  \end{cases}
  \label{eq:geometry_operator}
\end{equation}
for $q=0,\ldots,N_v-1$. Thus, each row of $\bPsi_{\mathcal G}$ contains at most two adjacent nonzero entries, whose locations and weights are determined by the corresponding physical lag $\tau_p$. The operator $\bPsi_{\mathcal G}$ is completely determined by the acquisition geometry $\mathcal G$ and the prescribed virtual-lag grid, and is therefore signal-independent.

Using $g_n=x_n+\varepsilon_n$, the single-look pairwise product in Eq.~\eqref{eq:pairwise_observation} can be written as
\begin{equation}
  y_p
  =
  r(\tau_p)
  +
  \varepsilon_{\mathrm{cross},p}
  +
  \varepsilon_{\mathrm{noise},p},
  \label{eq:pairwise_surrogate_model}
\end{equation}
where $\varepsilon_{\mathrm{noise},p} = x_{m_p}\varepsilon_{n_p}^{*} + \varepsilon_{m_p}x_{n_p}^{*} + \varepsilon_{m_p}\varepsilon_{n_p}^{*}$ collects the signal-noise and noise-noise perturbations. Collecting the pairwise observations as
$\by=[y_1,\ldots,y_P]^{\top}$ and combining
Eq.~\eqref{eq:pairwise_surrogate_model} with
Eq.~\eqref{eq:virtual_lag_interp} yields the structured
single-look lag-domain observation model
\begin{equation}
  \by
  =
  \bPsi_{\mathcal G}\bc
  +
  \boldsymbol{\varepsilon}_{\mathrm{c}},
  \label{eq:lag_forward_model}
\end{equation}
where $\boldsymbol{\varepsilon}_{\mathrm{c}} = \boldsymbol{\varepsilon}_{\mathrm{cross}} + \boldsymbol{\varepsilon}_{\mathrm{noise}} + \boldsymbol{\varepsilon}_{\mathrm{interp}}$ collects the multi-scatterer cross terms, propagated measurement noise, and virtual-lag interpolation errors.

Equation~\eqref{eq:lag_forward_model} therefore defines the single-look lag-domain observation model used in the subsequent structured estimation, with $\boldsymbol{\varepsilon}_{\mathrm c}$ accounting for cross-term, noise, and interpolation residuals. To provide a statistical interpretation of the desired lag response, consider the random-phase scatterer model $\gamma_k=A_k e^{j\varphi_k}$, where the phases $\varphi_k$ are mutually independent and uniformly distributed over $(-\pi,\pi]$. Under this model, the multi-scatterer cross terms vanish in expectation. For zero-mean measurement noise that is independent of the signal and across acquisitions, the noise-related terms also vanish in expectation for $m\neq n$. Consequently,
\begin{equation}
  \mathbb{E}\!\left[
    g_mg_n^{*}
  \right]
  =
  r(b_m-b_n),
  \qquad m\neq n,
  \label{eq:pairwise_expectation}
\end{equation}
so that $r(\tau)$ coincides with the ensemble covariance sequence under the random-phase model. For an individual single-look realization, the cross-term and noise residuals remain, motivating structured estimation of the latent sequence $\bc$.

The use of baseline differences as second-order sampling locations is related to the difference-coarray concept in sparse array processing \cite{Pal2010,Vaidyanathan2011}. In contrast to conventional multi-snapshot coarray processing, the present formulation operates on a single TomoSAR observation vector and therefore requires explicit structured estimation of the latent virtual-lag representation. Equation~\eqref{eq:lag_forward_model} provides the observation model for the structured lag-domain estimator developed next.

\subsection{Model-Based Proximal-Projection Splitting for Lag-Domain Estimation}

Given the structured single-look lag-domain model in Eq.~\eqref{eq:lag_forward_model}, the objective is to recover a latent lag sequence $\bc$ that is consistent with the geometry-dependent observations while retaining the Hermitian Toeplitz positive semidefinite structure established in Eq.~\eqref{eq:virtual_toeplitz_decomposition}. Defining the feasible set
\begin{equation}
  \mathcal{C}
  =
  \left\{
    \bZ\in\mathbb{C}^{N_v\times N_v}:
    \bZ=\bZ^{H},\;
    \bZ\succeq0,\;
    \bZ\ \text{is Toeplitz}
  \right\},
  \label{eq:constraint_set}
\end{equation}
the structured lag-domain estimation problem is formulated as
\begin{equation}
  \hat{\bc}
  =
  \arg\min_{\bc}\;
  \frac{1}{2}
  \left\|
    \by-\bPsi_{\mathcal G}\bc
  \right\|_2^2
  +
  \zeta\mathcal{R}(\bc)
  +
  \iota_{\mathcal C}\!\left(\bT(\bc)\right),
  \label{eq:lag_opt}
\end{equation}
where $\mathcal{R}(\bc)$ represents a prior on the latent virtual-lag sequence, $\zeta>0$ controls its strength, and
$\iota_{\mathcal C}(\cdot)$ denotes the indicator function of $\mathcal C$. The three terms in Eq.~\eqref{eq:lag_opt} encode consistency with the single-look lag-domain observations, prior regularization, and the admissible Toeplitz-PSD spectral structure, respectively.

Motivated by the composite structure of Eq.~\eqref{eq:lag_opt}, we construct a model-based proximal-projection splitting scheme that sequentially handles these components and serves as the computational prototype for subsequent deep unfolding. Starting from a lag-domain state $\bu^{(\ell)}\in\mathbb{C}^{N_v}$, the regularization step is written as
\begin{equation}
  \bv^{(\ell)}
  =
  \operatorname{prox}_{\kappa_\ell \mathcal{R}}
  \!\left(
    \bu^{(\ell -1)}
  \right),
  \label{eq:reg_update}
\end{equation}
where $\kappa_\ell>0$ denotes an effective proximal parameter incorporating the regularization strength and step size.

The regularized estimate is subsequently reconciled with the observed pairwise products through a geometry-dependent quadratic proximal step, which serves as an analytical data-consistency operation:
\begin{equation}
\begin{aligned}
 \bc^{(\ell)}
 &=
 \arg\min_{\bc}\;
 \frac{1}{2}
 \left\|
  \by-\bPsi_{\mathcal G}\bc
 \right\|_2^2
 +
 \frac{\rho_{\ell}}{2}
 \left\|
  \bc-\bv^{(\ell)}
 \right\|_2^2
 \\
 &=
 \left(
  \bPsi_{\mathcal G}^{H}\bPsi_{\mathcal G}
  +
  \rho_{\ell}\bI
 \right)^{-1}
 \left(
  \bPsi_{\mathcal G}^{H}\by
  +
  \rho_{\ell}\bv^{(\ell)}
 \right),
 \label{eq:dc_update}
\end{aligned}
\end{equation}
where $\rho_{\ell}>0$ controls the coupling between the observation-consistent estimate and the regularized state. The update is parameterized by the geometry-dependent operator $\bPsi_{\mathcal G}$ and returns an $N_v$-dimensional lag-domain estimate.

The resulting lag sequence is next mapped to its Hermitian Toeplitz matrix and projected toward the feasible set $\mathcal C$. Since $\mathcal C$ is the intersection of the Hermitian Toeplitz subspace and the
positive semidefinite cone, we employ Dykstra's alternating-projection algorithm \cite{Dykstra1983}. With $J$ Dykstra iterations, we denote the resulting approximation to the projection onto $\mathcal C$ by $\Pi_{\mathcal C}^{(J)}(\cdot)$, yielding
\begin{equation}
  \widetilde{\bT}^{(\ell)}
  =
  \Pi_{\mathcal C}^{(J)}
  \!\left(
    \bT(\bc^{(\ell)})
  \right).
  \label{eq:structured_matrix_update}
\end{equation}

To return the structured matrix estimate to the lag domain, we define the lag-extraction operator
$\mathcal{E}:\mathbb{C}^{N_v\times N_v}\rightarrow\mathbb{C}^{N_v}$ as
\begin{equation}
  \left[
    \mathcal{E}(\bZ)
  \right]_q
  =
  \frac{1}{N_v-q}
  \sum_{i=1}^{N_v-q}
  Z_{i+q,i},
  \qquad
  q=0,\ldots,N_v-1.
  \label{eq:lag_extraction}
\end{equation}
For an exact Hermitian Toeplitz matrix, this operation recovers its nonnegative-lag sequence. For a finite-step projection, diagonal averaging also reduces the residual departure from Toeplitz structure. The resulting lag sequence $\bu^{(\ell)}=\mathcal{E}(\widetilde{\bT}^{(\ell)})$ is then fed into the next iteration. This proximal-projection scheme provides the model-based backbone for the unfolded architecture developed next.

\subsection{Geometry-Decoupled Deep Unfolding}

\begin{figure*}[t]
    \centering
    \includegraphics[width=0.85\textwidth]{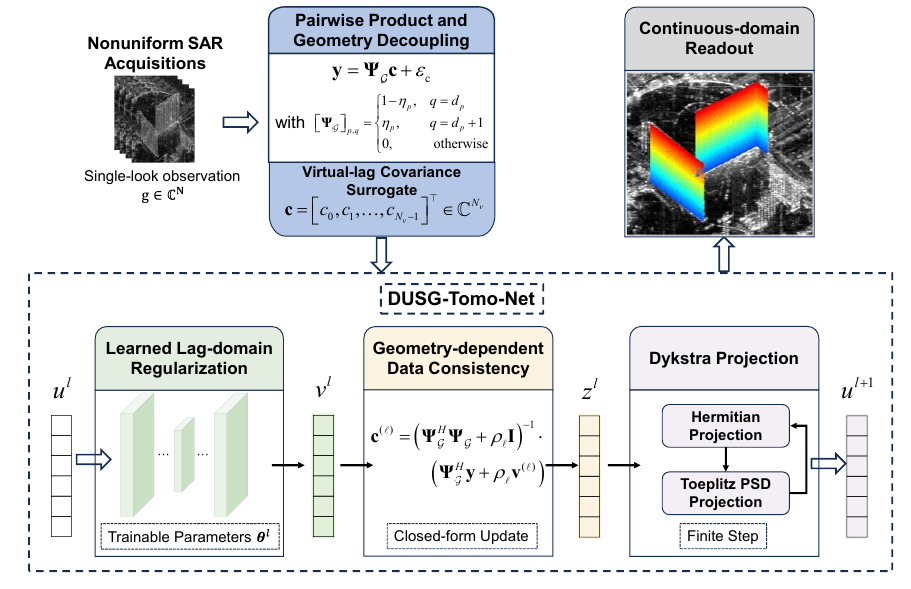}
    \caption{Overall architecture of DUSG-Tomo-Net. The acquisition-dependent geometry is incorporated analytically through $\bPsi_{\mathcal G}$, while the learned regularization operates on the common virtual-lag representation. Each unfolded layer combines learned regularization, geometry-dependent data consistency, and Toeplitz-PSD structural projection. Root-MUSIC processes the final structured estimate to obtain continuous-domain elevation estimates.}
    \label{fig:dusg_framework}
\end{figure*}

Based on the proximal-projection scheme developed in Section~III-B, we construct DUSG-Tomo-Net by unfolding the iterations into trainable layers. The overall architecture is illustrated in Fig.~\ref{fig:dusg_framework}. The analytical geometry-dependent data-consistency and structured-projection operations are retained, while the lag-domain proximal mapping is replaced by a learned regularization module operating on the common virtual-lag representation.

Let $\bu^{(\ell)}\in\mathbb{C}^{N_v}$ denote the lag-domain state at layer $\ell$, with $\bu^{(0)}$ given by the initialization and $\ell=1,\ldots,M$. Given the pairwise observation vector $\by$ and the geometry operator $\bPsi_{\mathcal G}$, the network is initialized by the analytical backprojection
\begin{equation}
  \bu^{(0)}
  =
  \bPsi_{\mathcal G}^{H}\by.
  \label{eq:initial_backprojection}
\end{equation}

The abstract proximal mapping in Eq.~\eqref{eq:reg_update} is replaced by a learned layer-dependent regularization module, with the effective proximal parameter absorbed into the trainable mapping. Specifically, we parameterize the regularization update in residual form as
\begin{equation}
  \bv^{(\ell)}
  =
  \mathcal{R}_{\theta_{\ell}}
  \!\left(
    \bu^{(\ell-1)}
  \right)
  =
  \bu^{(\ell-1)}
  +
  \mathcal{F}_{\theta_{\ell}}
  \!\left(
    \bu^{(\ell-1)}
  \right),
  \label{eq:learned_regularization}
\end{equation}
where $\mathcal{F}_{\theta_{\ell}}(\cdot)$ denotes the trainable residual mapping at the $\ell$th layer. The regularized state is then passed through the same model-based data-consistency and structural-projection steps derived in Section~III-B:
\begin{align}
  \bc^{(\ell)}
  &=
  \left(
    \bPsi_{\mathcal G}^{H}\bPsi_{\mathcal G}
    +
    \rho_{\ell}\bI
  \right)^{-1}
  \left(
    \bPsi_{\mathcal G}^{H}\by
    +
    \rho_{\ell}\bv^{(\ell)}
  \right),
  \label{eq:dc_layer}
  \\
  \widetilde{\bT}^{(\ell)}
  &=
  \Pi_{\mathcal C}^{(J)}
  \!\left(
    \bT(\bc^{(\ell)})
  \right),
  \label{eq:structure_layer}
  \\
  \bu^{(\ell)}
  &=
  \mathcal{E}
  \!\left(
    \widetilde{\bT}^{(\ell)}
  \right),
  \label{eq:lag_feedback}
\end{align}
where $\rho_{\ell}>0$ is a layer-dependent parameter controlling the balance between the learned regularized state and the observed lag-domain data.

The geometry-decoupled property follows directly from this decomposition. For an acquisition geometry $\mathcal G_i$, the number of admissible pairwise products $P_i$ may vary, leading to $\by_{\mathcal G_i}\in\mathbb{C}^{P_i}$ and
$\bPsi_{\mathcal G_i}\in\mathbb{R}^{P_i\times N_v}$. Despite these geometry-dependent input dimensions, the analytical
data-consistency step always maps the observations back to the common $N_v$-dimensional virtual-lag space. Consequently, the learned regularization modules $\mathcal{R}_{\theta_\ell}$ operate on a fixed-dimensional representation and contain no acquisition-specific sensing matrices or geometry-dependent weight matrices. The
geometry-specific information is therefore carried explicitly by $\bPsi_{\mathcal G}$, whereas the trainable parameters $\{\theta_\ell,\rho_\ell\}_{\ell=1}^{M}$ are shared across acquisition geometries.

This separation makes it possible to reuse the same trained unfolded model across different acquisition configurations by recomputing only the analytical geometry operator. However, such reuse requires the new physical baseline configuration to remain representable within the virtual-lag domain on which the trainable modules are defined. An acquisition geometry $\mathcal G$ is representationally compatible with the prescribed virtual-lag representation $(N_v,\Delta_b)$ if two conditions hold: its full physical baseline-difference range satisfies $B_{\max,\mathcal G} \leq (N_v-1)\Delta_b$, where $B_{\max,\mathcal G}\triangleq \max_{m,n}|b_m-b_n|$, and the prescribed elevation support satisfies the principal-branch condition specified in Section~III-A.

The physical baseline configuration still enters the initialization and every data-consistency layer through $\bPsi_{\mathcal G}$ and determines the available physical-lag coverage represented by that operator. Representational compatibility therefore enables the same network parameterization to be used after recomputing $\bPsi_{\mathcal G}$ but does not imply geometry-invariant reconstruction accuracy. Performance also depends on the physical-lag coverage and sampling density, the conditioning of $\bPsi_{\mathcal G}$, and the range of acquisition geometries represented during training.

In the implementation, each complex-valued lag sequence is represented by its real and imaginary components and processed by a compact residual CNN. The learned regularization modules use layer-specific trainable parameters
$\theta_{\ell}$, while the positive scalar $\rho_{\ell}$ is parameterized to remain strictly positive during training. The analytical operators $\bPsi_{\mathcal G}$ and $\bPsi_{\mathcal G}^{H}\bPsi_{\mathcal G}$ are constructed directly from the baseline geometry and are not trainable. After $M$ unfolded layers, DUSG-Tomo-Net outputs the final finite-step structured estimate $\widetilde{\bT}^{(M)}$, which is subsequently used for continuous-domain scatterer localization.

\subsection{Continuous-Domain Readout and Training Objective}

The final structured estimate $\widetilde{\bT}^{(M)}$ is converted into continuous-domain elevation estimates using Root-MUSIC \cite{root-music,Rao1989rootmusic}. For a prescribed or independently estimated model order $\widehat{K}$, we compute the eigendecomposition of $\widetilde{\bT}^{(M)}$ and construct the noise-subspace matrix $\widehat{\bU}_{\mathrm N}$ from the eigenvectors associated with its $N_v-\widehat{K}$ smallest eigenvalues. The Root-MUSIC polynomial is defined as
\begin{equation}
  q(z)
  =
  \ba_v^{T}(z^{-1})
  \widehat{\bU}_{\mathrm N}
  \widehat{\bU}_{\mathrm N}^{H}
  \ba_v(z),
  \label{eq:rootmusic_poly}
\end{equation}
with $\ba_v(z)=[1,z,\ldots,z^{N_v-1}]^{\top}$. Multiplication by $z^{N_v-1}$ converts $q(z)$ into an ordinary polynomial whose roots occur in conjugate-reciprocal pairs. From each relevant pair, the root inside the unit circle and closest to it is selected. Let $\{\hat z_k\}_{k=1}^{\widehat K}$ denote the selected roots. According to the virtual steering relation in Eq.~\eqref{eq:virtual_steering}, the corresponding continuous elevation estimates are obtained as
\begin{equation}
  \hat s_k
  =
  \frac{\lambda r}{4\pi\Delta_b}
  \arg(\hat z_k),
  \qquad
  k=1,\ldots,\widehat K.
  \label{eq:rootmusic_height_map}
\end{equation}
Under the principal-branch elevation-support condition specified in Section~III-A, the phase-to-elevation mapping in
Equation~\eqref{eq:rootmusic_height_map} is unambiguous over the prescribed elevation interval. For elevation supports extending beyond this interval, additional branch selection would be required. Since the final spectral localization is performed through polynomial rooting rather than searching over a discretized elevation dictionary, the proposed framework retains continuous-domain, off-grid elevation estimation. The model order $\widehat K$ required at inference is determined by the screening-and-validation procedure described in Section~V-C.

DUSG-Tomo-Net is trained using losses defined on the recovered structured representation, without differentiating through the subsequent polynomial root-selection procedure. The training objective is
\begin{equation}
  \mathcal{L}
  =
  \beta_1\mathcal{L}_{\mathrm{sub}}
  +
  \beta_2\mathcal{L}_{\mathrm{data}}
  +
  \beta_3\mathcal{L}_{\mathrm{tpl}},
  \label{eq:total_loss}
\end{equation}
where $\beta_1$, $\beta_2$, and $\beta_3$ control the relative contributions of the subspace, data-consistency, and structural terms, respectively.

The ground-truth model order $K$ is used only to construct the supervised subspace target during training and is not provided as an input to DUSG-Tomo-Net. For a training sample containing $K$ ground-truth scatterers, let $\bA_{\mathrm{gt}} = \left[ \ba_v(s_1),\ldots,\ba_v(s_K) \right]$ denote the corresponding virtual steering matrix. The orthogonal projector onto its noise subspace is then
\begin{equation}
  \bP_{\mathrm N}^{\mathrm{gt}}
  =
  \bI
  -
  \bA_{\mathrm{gt}}
  \bA_{\mathrm{gt}}^{\dagger},
  \label{eq:gt_noise_projector}
\end{equation}
where $(\cdot)^{\dagger}$ denotes the Moore-Penrose pseudoinverse. The subspace loss is defined as
\begin{equation}
  \mathcal{L}_{\mathrm{sub}}
  =
  \frac{
    \left\|
      \widetilde{\bT}^{(M)}
      \bP_{\mathrm N}^{\mathrm{gt}}
    \right\|_{F}^{2}
  }{
    \left\|
      \widetilde{\bT}^{(M)}
    \right\|_{F}^{2}
  }.
  \label{eq:subspace_loss}
\end{equation}
This term penalizes the energy of the recovered structured representation that leaks into the ground-truth noise subspace, thereby encouraging its range space to remain consistent with the ground-truth steering subspace. The data-consistency term is defined as
\begin{equation}
  \mathcal{L}_{\mathrm{data}}
  =
  \frac{
    \left\|
      \bPsi_{\mathcal G}
      \mathcal{E}
      \!\left(
        \widetilde{\bT}^{(M)}
      \right)
      -
      \by
    \right\|_{2}^{2}
  }{
    \|\by\|_{2}^{2}
  }.
  \label{eq:data_loss}
\end{equation}
The data-consistency loss constrains the recovered virtual-lag representation to remain consistent with the geometry-dependent single-look observations in Eq.~\eqref{eq:lag_forward_model}.

Although each unfolded layer incorporates Toeplitz-PSD structural enforcement, only a finite number of Dykstra iterations is performed. Since each Dykstra iteration terminates with projection onto the positive semidefinite cone, the resulting matrix remains positive semidefinite, while a small residual departure from exact Toeplitz structure may persist. We therefore introduce the structural loss:
\begin{equation}
  \mathcal{L}_{\mathrm{tpl}}
  =
  \frac{
    \left\|
      \widetilde{\bT}^{(M)}
      -
      \mathcal{P}_{\mathrm{Tpl}}
      \!\left(
        \widetilde{\bT}^{(M)}
      \right)
    \right\|_{F}^{2}
  }{
    \left\|
      \widetilde{\bT}^{(M)}
    \right\|_{F}^{2}
  },
  \label{eq:toeplitz_loss}
\end{equation}
where $\mathcal{P}_{\mathrm{Tpl}}(\cdot)$ denotes projection onto the Hermitian Toeplitz subspace. Reducing this residual improves consistency with the Toeplitz virtual-aperture model underlying the subsequent subspace-based spectral readout. Together, the three loss terms encourage the network output to preserve the ground-truth spectral subspace, remain consistent with the geometry-dependent single-look observations, and satisfy the required structured representation. The trainable network parameters are optimized through Eq.~\eqref{eq:total_loss}, whereas the geometry operator $\bPsi_{\mathcal G}$ and the structural operators remain analytically specified and nontrainable.

\section{Experimental Results}

\subsection{Experimental Setup}

\subsubsection{Simulation and Training Configuration}

The simulations are conducted under an X-band spaceborne TomoSAR configuration. A nominal aperture consisting of $N=20$ acquisitions is defined by uniformly spaced perpendicular baselines over $[-150,150]$~m, corresponding to a baseline span of $B=300$~m. With the radar parameters listed in Table~\ref{tab:radar_params}, the corresponding Rayleigh elevation resolution is $\rho_s=\lambda r/(2B)\approx36.17$~m.
\begin{table}[t]
\centering
\caption{Radar and aperture parameters used in the simulations.}
\label{tab:radar_params}
\begingroup
\footnotesize
\renewcommand{\arraystretch}{1.2}
\begin{tabular}{lll}
\toprule
\textbf{Parameter} & \textbf{Symbol} & \textbf{Value} \\
\midrule
Wavelength & $\lambda$ & $0.031$ m \\
Slant range & $r$ & $7.0\times10^{5}$ m \\
Nominal baseline span & $B$ & $300$ m \\
Number of acquisitions & $N$ & $20$ \\
Rayleigh elevation resolution & $\rho_s$ & $\approx36.17$ m \\
\bottomrule
\end{tabular}
\endgroup
\end{table}

A nonuniform training geometry $\mathcal{G}_{\mathrm{tr}}$ is generated from the nominal aperture by perturbing the interior baselines with independent zero-mean Gaussian offsets of standard deviation $10$~m. The two endpoint baselines remain fixed so that the baseline span is preserved at $300$~m, and perturbed interior baselines falling outside the prescribed aperture are projected back onto the interval $[-150,150]$~m. This geometry is used to generate the synthetic training and validation observations according to the TomoSAR signal model in Section~II-A.

The training set contains equal proportions of single- and double-scatterer samples. For a single-scatterer sample, the elevation is drawn continuously as $s \sim U(-150,150)$~m. The complex scattering coefficient is generated as $\gamma=A\exp(j\phi)$, with $A\sim{U}(1,4)$ and $\phi\sim{U}(-\pi,\pi]$. For a double-scatterer sample, the two components are generated independently according to the same elevation, amplitude, and phase distributions, thereby covering a broad range of scatterer separations, amplitude ratios, and phase differences. The SNR is sampled uniformly from $[0,15]$~dB.

DUSG-Tomo-Net is implemented in PyTorch \cite{pytorch}. Unless otherwise stated, the virtual-lag representation uses $N_v=32$ samples with $\Delta_b=10$~m, and the network contains $M=12$ unfolded layers. Each learned regularization block consists of three one-dimensional convolutional layers with a kernel size of $5$ and $16$ hidden channels, and $J=3$ Dykstra iterations are used for the structured projection in each unfolded layer. The maximum model order considered during training is $K_{\max}=2$.

The training and validation sets contain $100\,000$ and $20\,000$ synthetic pixels, respectively. The network is trained for $100$ epochs using the Adam optimizer \cite{adam}, with an initial learning rate of $10^{-3}$, a batch size of $1000$, and cosine-annealing learning-rate scheduling. The loss weights in Eq.~\eqref{eq:total_loss} are fixed to $\beta_1=1.0$, $\beta_2=0.5$, and $\beta_3=0.1$.

\subsubsection{Comparison Methods and Evaluation Protocol}

The proposed DUSG-Tomo-Net is compared with representative TomoSAR reconstruction methods spanning different reconstruction paradigms and elevation-resolution capabilities. SVD-Wiener is included as a conventional reference, whereas SL1MMER \cite{Zhu2011tomo} and $\gamma$-Net \cite{Qian2022gammanet} represent grid-based sparse super-resolution methods. TADCG \cite{Shao2024TADCG} is adopted as a representative gridless method designed for nonuniform baselines. For methods that rely on a discretized elevation dictionary, the grid spacing is set to $1$~m unless otherwise specified. All methods are evaluated using the same physical baseline configuration and noisy TomoSAR observations at each operating point. Elevation accuracy is quantified using the root-mean-square error (RMSE). For double-scatterer cases, the estimated and reference elevations are optimally paired before the localization error is calculated. The effective detection rate defined in \cite{Qian2022gammanet} is used to evaluate super-resolution capability. The specific test geometries, SNR levels, and Monte Carlo settings are defined in the corresponding experimental subsections.

\subsection{Continuous-Domain Localization}

We first evaluate the continuous-domain elevation localization capability of the proposed method using single-scatterer simulations. A total of 1000 independent samples are generated, with the scatterer elevation drawn continuously and uniformly from $[-20,20]$~m. The SNR is fixed at 6~dB, and the nonuniform-baseline configuration described in Section~IV-A is used. For reconstruction methods relying on a discretized elevation dictionary, the elevation grid spacing is set to $1$~m. DUSG-Tomo-Net is compared with TADCG, $\gamma$-Net, SL1MMER, and SVD-Wiener.

\begin{figure*}[t]
\centering
\includegraphics[width=\textwidth]{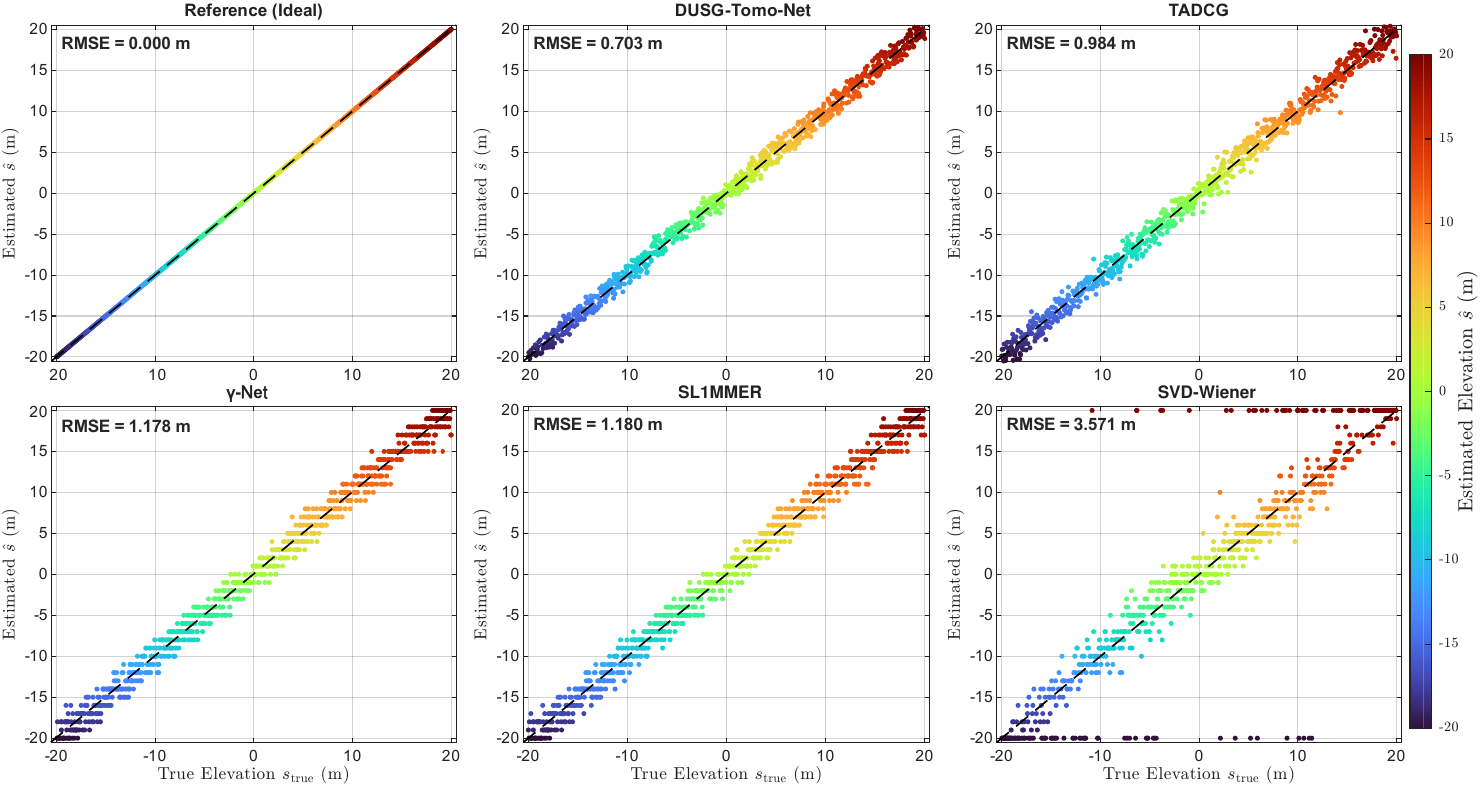}
\caption{Continuous-domain elevation localization for single-scatterer scenes under nonuniform baselines ($K=1$, SNR $=6$~dB, $N=20$). The scatterer elevation is drawn continuously and uniformly from $[-20,20]$~m. The dashed diagonal denotes ideal localization, and the RMSE of each method is reported in the corresponding panel. For methods based on discretized elevation sampling, the elevation grid spacing is $1$~m.}
\label{fig:continuous_localization}
\end{figure*}

Figure~\ref{fig:continuous_localization} compares the estimated elevations with their reference values. DUSG-Tomo-Net and TADCG produce continuously varying estimates distributed around the ideal diagonal, whereas $\gamma$-Net and SL1MMER exhibit clearly visible staircase patterns because their estimates are constrained to the predefined elevation grid. The staircase behavior remains evident even with a $1$~m grid spacing, which is much finer than the nominal Rayleigh resolution. This result illustrates the distinction between refining a discretized reconstruction grid and estimating scatterer elevations directly in the continuous domain.

Among the compared methods, DUSG-Tomo-Net provides the most concentrated estimates around the ideal diagonal and achieves the lowest localization RMSE of $0.703$~m. TADCG, which also performs continuous-domain estimation under nonuniform baselines, achieves an RMSE of $0.984$~m but exhibits noticeably greater dispersion. The grid-based $\gamma$-Net and SL1MMER obtain similar RMSEs of $1.178$~m and $1.180$~m, respectively, while showing the characteristic staircase behavior imposed by the discretized elevation grid. In contrast, SVD-Wiener yields a substantially larger RMSE of $3.571$~m and produces a considerable number of anomalous estimates, particularly near the boundaries of the elevation search interval. This behavior is mainly attributed to the sidelobe ambiguity of conventional spectral reconstruction, in which sidelobe peaks may be mistaken for the dominant scatterer under noisy, finite-aperture observations. These results confirm the continuous-domain localization capability of DUSG-Tomo-Net under nonuniform baseline sampling.

\subsection{Super-Resolution Performance}

We further evaluate the super-resolution capability of DUSG-Tomo-Net using double-scatterer simulations under the nonuniform-baseline configuration described in Section~IV-A. Two scatterers are placed with normalized separation
$\alpha=d_s/\rho_s\in\{0.2,0.3,\ldots,1.2\}$, where $d_s$ is the elevation spacing and $\rho_s$ denotes the Rayleigh elevation resolution. The SNR is set to 0, 6, and 10~dB, and 2000 Monte Carlo trials are conducted at each operating point. We evaluate super-resolution performance using the effective detection rate defined in \cite{Qian2022gammanet}. This criterion is stricter than simply reporting the presence of two peaks, because it requires the two components to be simultaneously separated and accurately localized. DUSG-Tomo-Net is compared with $\gamma$-Net, SL1MMER, and TADCG.

\begin{figure*}[t]
\centering
\includegraphics[width=0.98\textwidth]{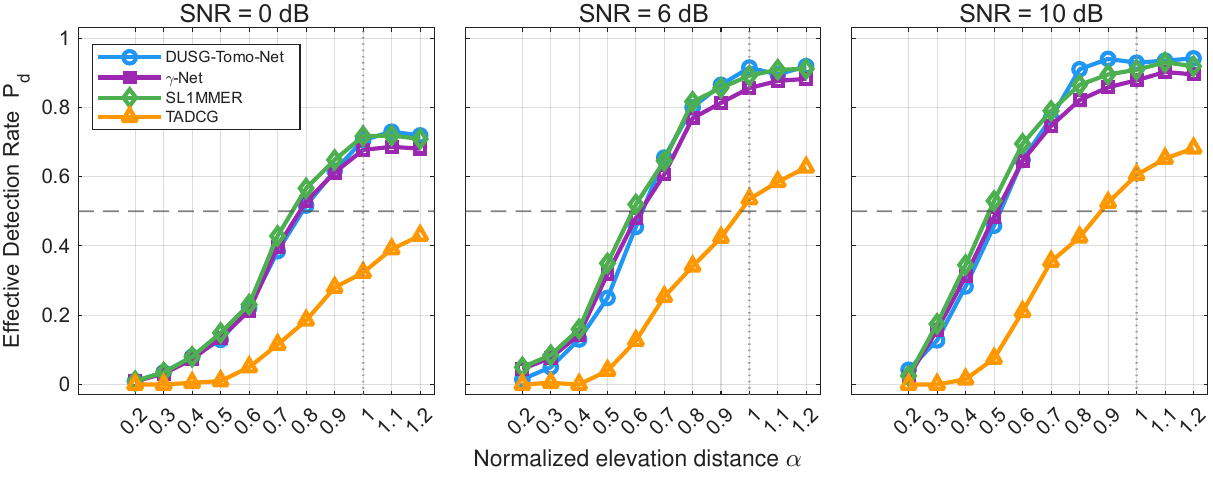}
\caption{Effective detection rate versus normalized scatterer separation $\alpha=d_s/\rho_s$ for the double-scatterer super-resolution experiment under nonuniform baselines. Results are shown for SNRs of 0, 6, and 10~dB.}
\label{fig:superres_pd}
\end{figure*}

As shown in Fig.~\ref{fig:superres_pd}, the separation performance of all methods improves with increasing $\alpha$ and SNR. DUSG-Tomo-Net exhibits a rapid transition to successful two-scatterer separation in the sub-Rayleigh regime. The effective detection rate reaches approximately 0.5 at $\alpha\approx0.8$, $0.6$, and $0.5$ for SNRs of 0, 6, and 10~dB, respectively. At the Rayleigh limit, DUSG-Tomo-Net achieves effective detection rates of approximately 0.72, 0.90, and 0.93, whereas TADCG reaches only about 0.33, 0.54, and 0.61, respectively. $\gamma$-Net and SL1MMER remain competitive, particularly at moderate and high SNR, and DUSG-Tomo-Net achieves comparable super-resolution performance.

\begin{figure*}[t]
\centering
\includegraphics[width=\textwidth]{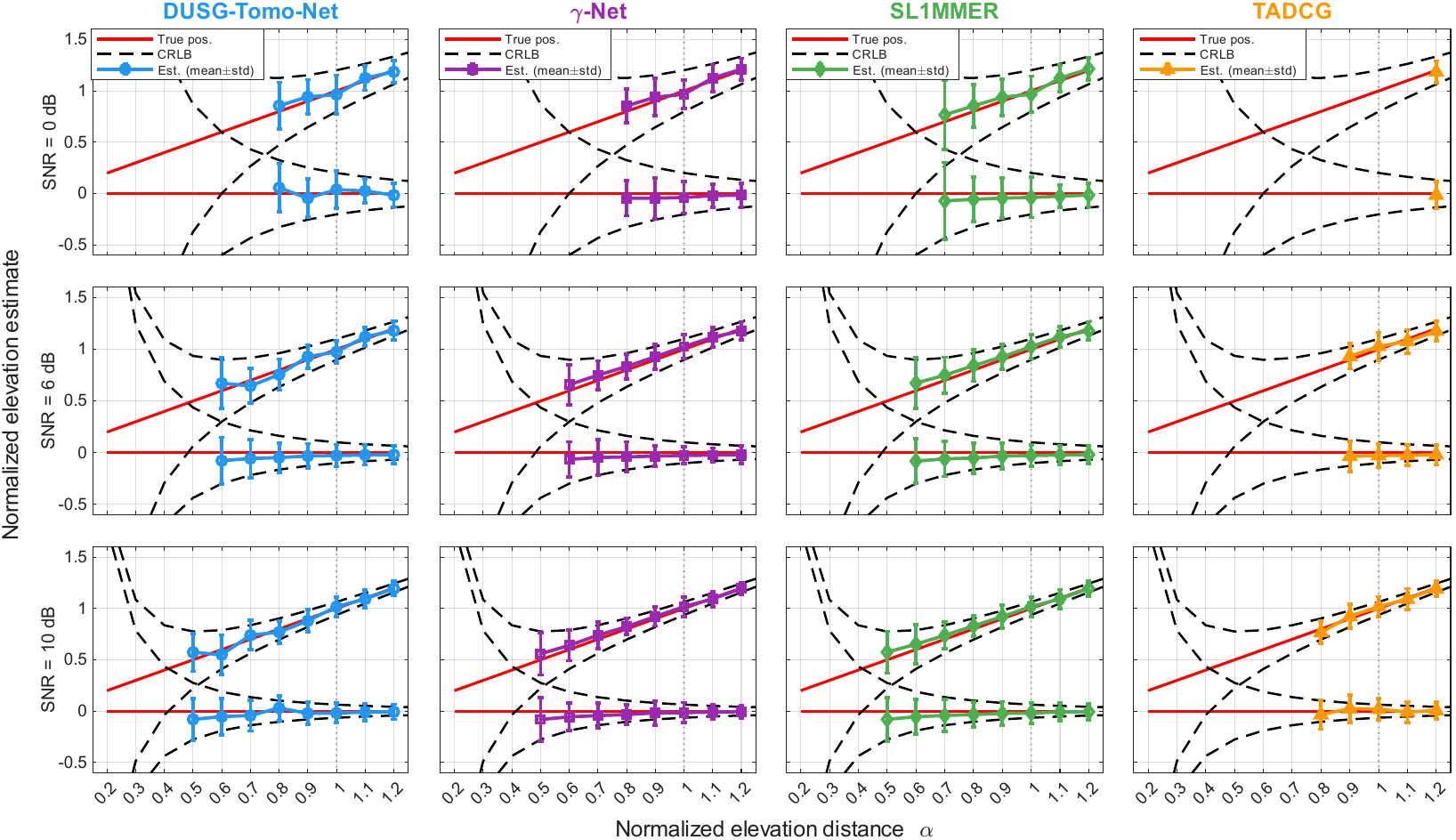}
\caption{Normalized elevation estimation results for the double-scatterer super-resolution experiment. Red solid lines denote the true lower and upper scatterer positions, black dashed lines indicate the CRLB envelopes, and colored error bars show the mean $\pm$ standard deviation of the estimated elevations. Only operating points with $P_d \geqslant 0.4$ are included to ensure that the reported statistics are supported by a sufficient number of successful double-scatterer reconstructions.}
\label{fig:superres_accuracy}
\end{figure*}

Figure~\ref{fig:superres_accuracy} provides a complementary view by showing the mean and standard deviation of the recovered scatterer elevations. For statistical reliability, only operating points with an effective detection rate greater than 0.4 are included. DUSG-Tomo-Net tracks both elevation branches with little bias, while the estimation variance decreases with increasing SNR and scatterer spacing. At 6 and 10~dB, stable two-branch reconstruction is observed well below the Rayleigh limit, and the estimates progressively approach the CRLB as $\alpha$ increases. In contrast, TADCG requires noticeably greater scatterer spacing before stable two-branch reconstruction emerges, consistent with its lower effective detection rate in Fig.~\ref{fig:superres_pd}. $\gamma$-Net and SL1MMER show similarly stable behavior in their resolvable regimes. Overall, DUSG-Tomo-Net achieves reliable near- and sub-Rayleigh separation under nonuniform baselines, with performance comparable to the strong grid-based baselines and clearly better than that of TADCG.


\subsection{Generalization Across Acquisition Geometries}

A key feature of DUSG-Tomo-Net is the separation of the geometry-dependent measurement mapping from the trainable reconstruction parameters. We therefore evaluate the reuse of a single trained model under varying acquisition geometries. Throughout this subsection, all learned parameters are kept fixed, while the geometry-dependent operator $\bPsi_{\mathcal G}$ is recomputed from the actual baseline configuration of each test geometry. Two complementary variations are considered: perturbations of the baseline positions with a fixed number of acquisitions and changes in the number of acquisitions. These experiments assess sensitivity to changes in the baseline configuration and adaptability to different acquisition numbers, respectively.


\subsubsection{Generalization to Perturbed Nonuniform Baselines}

We first examine the effect of changes in the physical baseline configuration while keeping the number of acquisitions fixed at $N=20$. Starting from the reference nonuniform geometry used for training, additional independent Gaussian perturbations are applied to the interior baseline positions, while the two aperture endpoints remain fixed. The perturbation standard deviation is varied as $\sigma_b \in \{0,2,5,8,10,15,20\}~\mathrm{m}$, where $\sigma_b=0$ corresponds to the reference training geometry and increasing $\sigma_b$ produces progressively larger deviations from it.

To account for the variability among individual baseline realizations, 20 independent acquisition geometries are generated at each perturbation level. Monte Carlo simulations are then performed independently for each geometry. The experiment is conducted at an SNR of 10~dB using a challenging double-scatterer configuration with normalized separation $\alpha=0.6$, which lies in the sub-Rayleigh transition regime identified in Section~IV-C. Performance is evaluated in terms of the effective detection rate and elevation RMSE. For the latter, the RMSE is first computed separately for each baseline geometry over the corresponding Monte Carlo trials and is subsequently averaged over the 20 geometry realizations. The reported standard deviation characterizes the geometry-to-geometry variability at each perturbation level.

DUSG-Tomo-Net is compared with $\gamma$-Net, SL1MMER, and TADCG. For SL1MMER and TADCG, the sensing model is constructed using the actual physical baselines of each test geometry. In contrast, $\gamma$-Net retains the parameters learned for the reference geometry and is applied directly to the perturbed configurations without retraining or fine-tuning.

\begin{figure*}[t]
\centering
\includegraphics[width=0.9\textwidth]{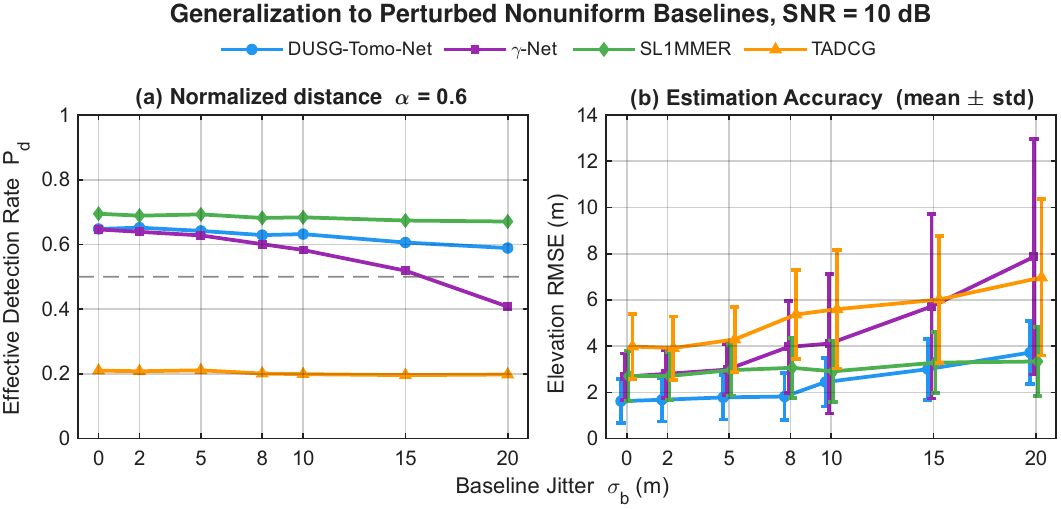}
\caption{Generalization performance under progressively perturbed nonuniform baseline configurations at SNR $=10$~dB and $\alpha=0.6$. For each perturbation level, 20 independent baseline geometries are generated and evaluated using Monte Carlo simulations. The horizontal axis denotes the standard deviation $\sigma_b$ of the additional Gaussian perturbation relative to the reference nonuniform training geometry. (a) Mean effective detection rate. (b) Mean elevation RMSE averaged over the 20 baseline geometries, with error bars indicating the standard deviation across geometry realizations.}
\label{fig:baseline_generalization}
\end{figure*}

As shown in Fig.~\ref{fig:baseline_generalization}(a), DUSG-Tomo-Net maintains relatively stable separation performance as the physical baselines increasingly deviate from the training configuration. Its effective detection rate decreases slightly from about 0.65 at $\sigma_b=0$ to about 0.6 at $\sigma_b=20$~m. In contrast, $\gamma$-Net exhibits a substantially stronger dependence on the baseline perturbation. Although its effective detection rate is comparable to that of DUSG-Tomo-Net near the reference geometry, it decreases progressively as $\sigma_b$ increases and reaches approximately 0.41 at the largest perturbation level. The difference becomes particularly pronounced for large geometry deviations, where the fixed geometry-coupled parameters of $\gamma$-Net are increasingly mismatched with the physical sensing configuration. SL1MMER remains stable over the tested perturbation range and provides a slightly higher effective detection rate than DUSG-Tomo-Net. TADCG also exhibits limited variation with increasing perturbation, although its effective detection rate remains substantially lower under this challenging sub-Rayleigh condition.

Figure~\ref{fig:baseline_generalization}(b) further evaluates localization accuracy and its sensitivity to individual baseline realizations. DUSG-Tomo-Net achieves the lowest mean RMSE over most of the perturbation range and exhibits relatively limited geometry-to-geometry variation. Its mean error increases gradually as $\sigma_b$ becomes larger, reflecting the influence of the changed physical sampling configurations. In contrast, $\gamma$-Net shows both a stronger increase in mean RMSE and a pronounced growth in variability across the independently generated geometries, particularly at large perturbation levels. TADCG likewise exhibits increasing localization error and geometry-to-geometry dispersion, indicating that different nonuniform sampling configurations can themselves lead to different levels of inversion difficulty. SL1MMER remains comparatively stable and approaches the RMSE of DUSG-Tomo-Net at the largest perturbation levels.


\subsubsection{Generalization to Different Numbers of Acquisitions}

We next examine the effect of varying the number of acquisitions. The reference configuration contains $N=20$ acquisitions, while the number of test acquisitions varies over $N \in \{8,12,16,20,24,28\}$. For $N<20$, the two endpoint baselines of the reference aperture are retained and the remaining $N-2$ acquisitions are randomly selected from the interior baseline positions of the reference nonuniform configuration. For $N>20$, the reference aperture is retained and additional baseline positions are randomly inserted within the same baseline range. In all cases, the two aperture endpoints remain fixed so that the total baseline span and corresponding nominal Rayleigh elevation resolution remain unchanged. For each value of $N$, 20 independent acquisition geometries are generated. Monte Carlo simulations are then performed separately for each geometry under the double-scatterer setup with SNR $=10$~dB and normalized separation $\alpha=0.6$.

\begin{figure*}[t]
\centering
\includegraphics[width=0.9\textwidth]{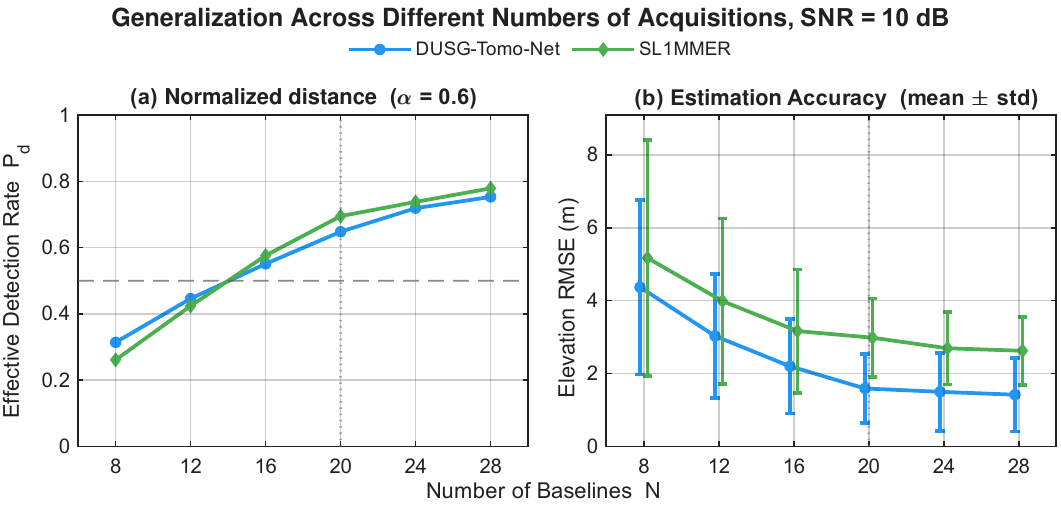}
\caption{Generalization performance for different numbers of acquisitions under a fixed baseline aperture (SNR $=10$~dB, $K=2$, and $\alpha=0.6$). For each $N$, 20 independent nonuniform acquisition geometries are generated and evaluated using Monte Carlo simulations. (a) Mean effective detection rate. (b) Mean elevation RMSE, with error bars indicating the standard deviation across geometry realizations.}
\label{fig:varying_acquisitions}
\end{figure*}

The experimental results in Fig.~\ref{fig:varying_acquisitions} demonstrate that DUSG-Tomo-Net maintains performance comparable to that of the learning-free SL1MMER algorithm across all tested acquisition numbers, although the network was trained only with $N=20$. As shown in Fig.~\ref{fig:varying_acquisitions}(a), the effective detection rate increases with the number of acquisitions for both methods, reflecting the additional measurement information provided by denser baseline sampling. DUSG-Tomo-Net achieves an effective detection rate of approximately 0.32 with only eight acquisitions, increasing progressively to about 0.76 at $N=28$. Its super-resolution performance remains comparable to that of SL1MMER over the entire range. The elevation estimation results in Fig.~\ref{fig:varying_acquisitions}(b) show a clearer advantage in localization accuracy. The mean RMSE of DUSG-Tomo-Net decreases from approximately $4.5$~m at $N=8$ to about $1.4$~m at $N=28$ and remains consistently below that of SL1MMER for all tested acquisition numbers.

Because SL1MMER explicitly reconstructs the steering matrix for each acquisition configuration, it provides a geometry-matched model-based reference that is unaffected by cross-geometry learning mismatch. The comparable performance of DUSG-Tomo-Net across the tested acquisition numbers therefore supports the effectiveness of the proposed geometry-decoupled formulation in accommodating changes in the physical acquisition geometry.

\subsection{Real-data Experiments}

\begin{table}[t]
\centering
\caption{CH-1 system parameters.}
\label{tab:CH1_params}
\begingroup
\footnotesize
\renewcommand{\arraystretch}{1.2}
\begin{tabular}{lll}
\toprule
\textbf{Parameter} & \textbf{Symbol} & \textbf{Value} \\
\midrule
Wavelength & $\lambda$ & $0.056$ m \\
Slant range & $r$ & $6.89\times10^{5}$ m \\
Incidence angle & $\theta$ & $41.01^{\circ}$ \\
Range resolution & $\rho_r$ & $0.6$ m \\
Azimuth resolution & $\rho_a$ & $1.2$ m \\
Baseline span & $B$ & $363$ m \\
Number of acquisitions & $N$ & $16$ \\
Rayleigh elevation resolution & $\rho_s$ & $\approx52.77$ m \\
\bottomrule
\end{tabular}
\endgroup
\end{table}

\begin{figure}[t]
\centering
\includegraphics[width=0.95\columnwidth]{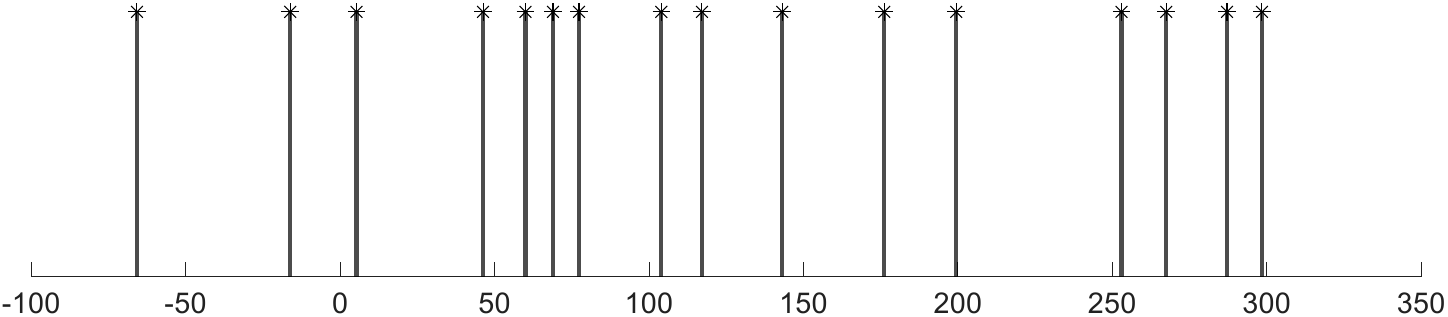}
\caption{Perpendicular-baseline configuration of the 16 CH-1 acquisitions used in the real-data experiment.}
\label{fig:baseline_distribution}
\end{figure}

\begin{figure}[t]
\centering
\includegraphics[width=0.95\columnwidth]{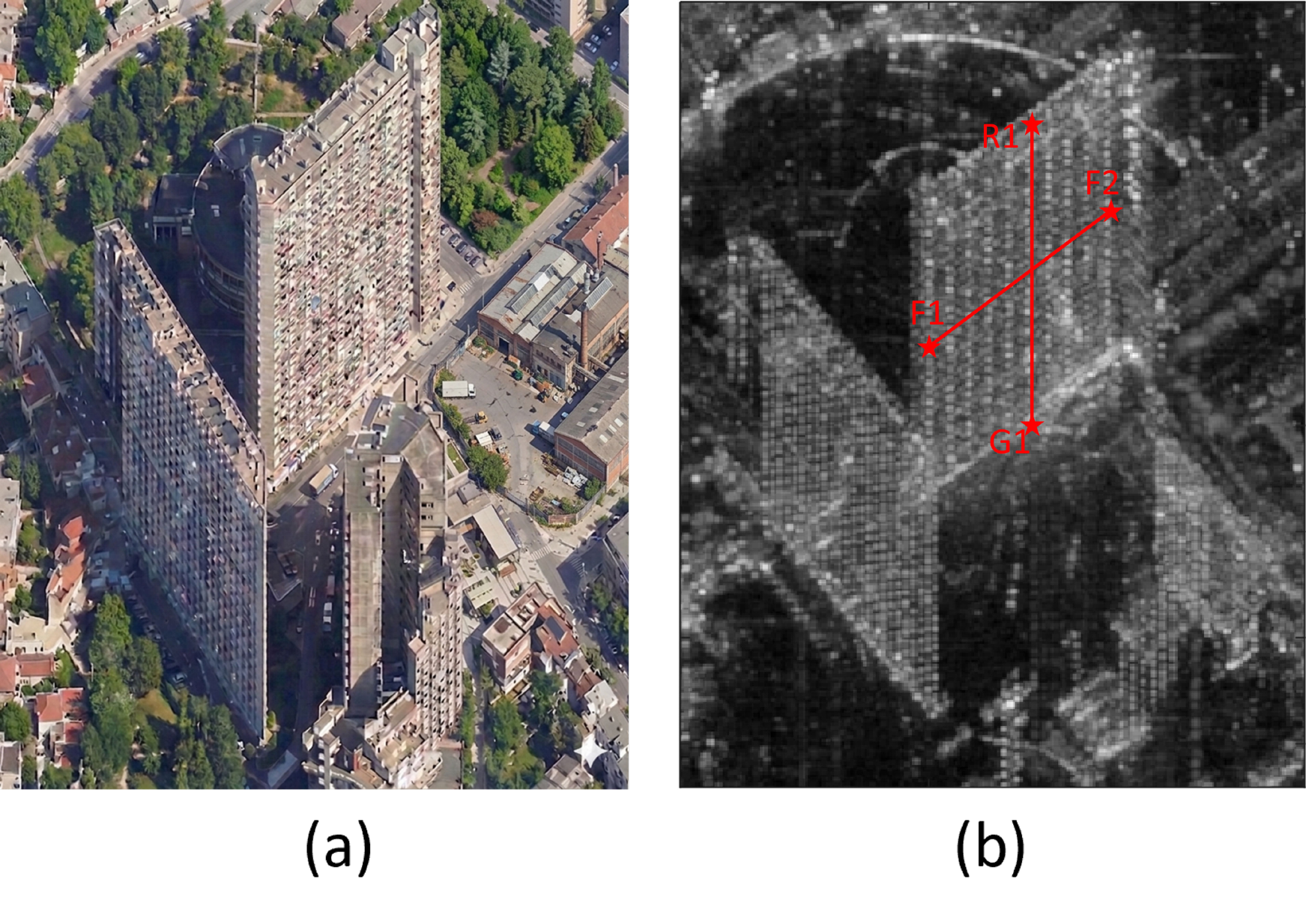}
\caption{Real-data test area. Left: UAV optical image. Right: SAR mean-intensity image. The selected scene contains two high-rise buildings with pronounced facade-related layover.}
\label{fig:real_test_area}
\end{figure}

To evaluate DUSG-Tomo-Net on real SAR data, we use a stack of 16 high-resolution staring-spotlight images acquired by CH-1. The images have spatial resolutions of 0.6~m in range and 1.2~m in azimuth. As shown in Fig.~\ref{fig:baseline_distribution}, the perpendicular baselines range from $-65$~m to $298$~m, corresponding to a baseline span of approximately $363$~m and a Rayleigh elevation resolution of $52.77$~m. The relevant CH-1 parameters are listed in Table~\ref{tab:CH1_params}. For the real-data experiment, a separate DUSG-Tomo-Net model is trained for the CH-1 configuration using $N_v=42$ and $\Delta_b=10$~m, following the synthetic training protocol in Section~IV-A. The model trained with the full 16-acquisition reference configuration is then kept fixed for the $N=12$ and $N=8$
subsets, with only $\bPsi_{\mathcal G}$ recomputed. Figure~\ref{fig:real_test_area} shows the UAV optical image and the SAR mean-intensity image of the test area. The selected region of interest (ROI) contains two high-rise buildings with pronounced facade-related layover and double-scattering responses. Because no external 3-D reference is available, the real-data analysis focuses on the spatial consistency of the reconstructed elevation structures and the separation characteristics of the detected double-scatterer responses.

\begin{figure*}[t]
\centering
\includegraphics[width=0.85\linewidth]{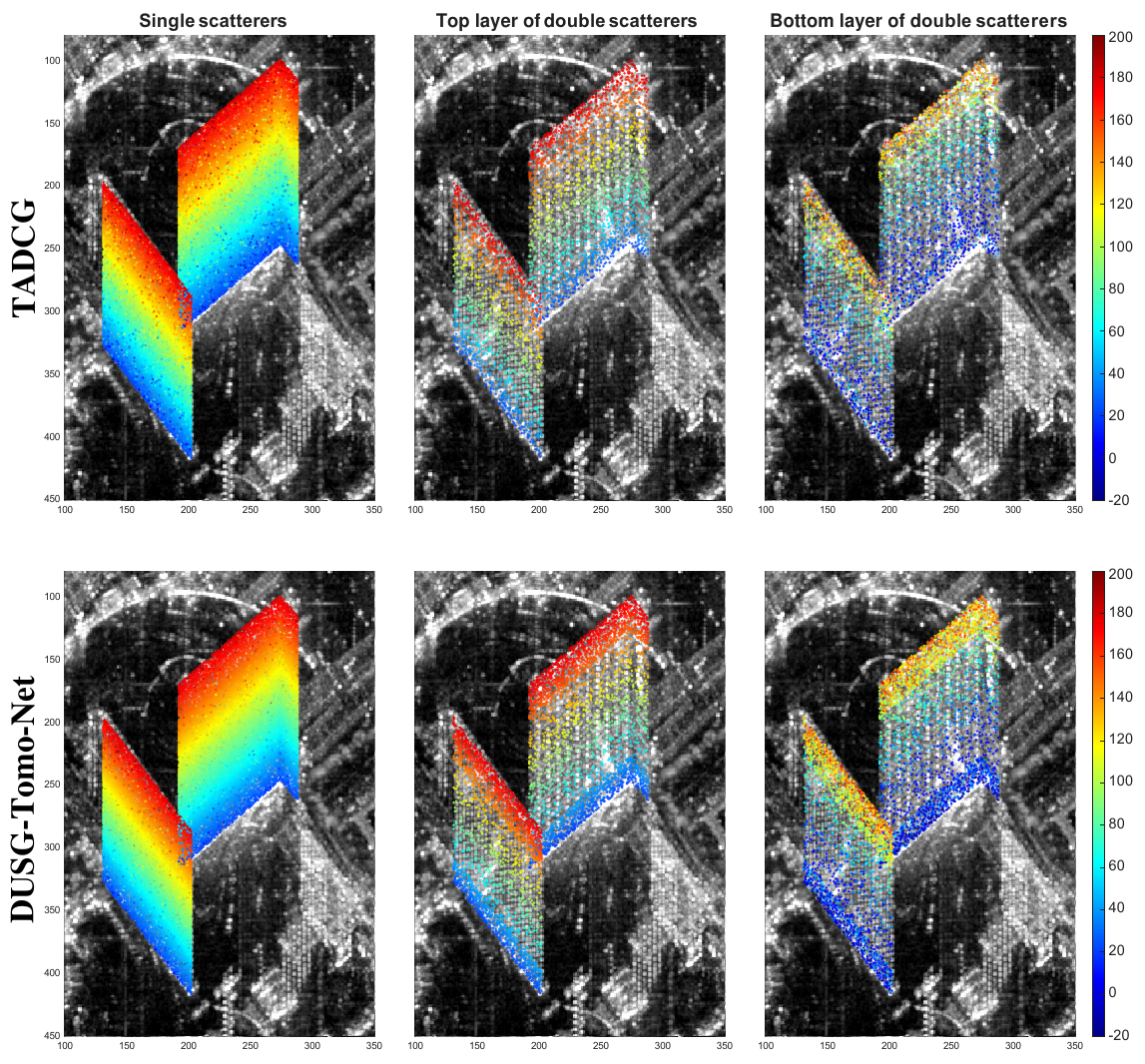}
\caption{Color-coded elevation reconstruction of the real scene using TADCG and DUSG-Tomo-Net with the full 16-image stack. The top and bottom rows show the TADCG and DUSG-Tomo-Net results, respectively. From left to right, the columns correspond to the detected single-scatterer layer and the upper and lower layers of the detected double-scatterer pixels.}
\label{fig:real_reconstruction_comparison}
\end{figure*}

Figure~\ref{fig:real_reconstruction_comparison} compares the elevation reconstructions obtained from the full 16-image stack. Both methods recover the dominant building geometry and the overall elevation variation along the two facades. The single-scatterer estimates produced by DUSG-Tomo-Net are spatially coherent and closely follow the major facade structures. More noticeable differences appear in the double-scatterer results. DUSG-Tomo-Net produces more continuous upper- and lower-layer responses over the layover-dominated facade regions, whereas the corresponding TADCG reconstructions contain more local gaps and fragmented detections. In particular, the lower double-scatterer layer reconstructed by DUSG-Tomo-Net forms more persistent structures over both buildings, indicating more stable separation of closely spaced elevation components.

\begin{figure*}[t]
\centering
\includegraphics[width=0.85\linewidth]{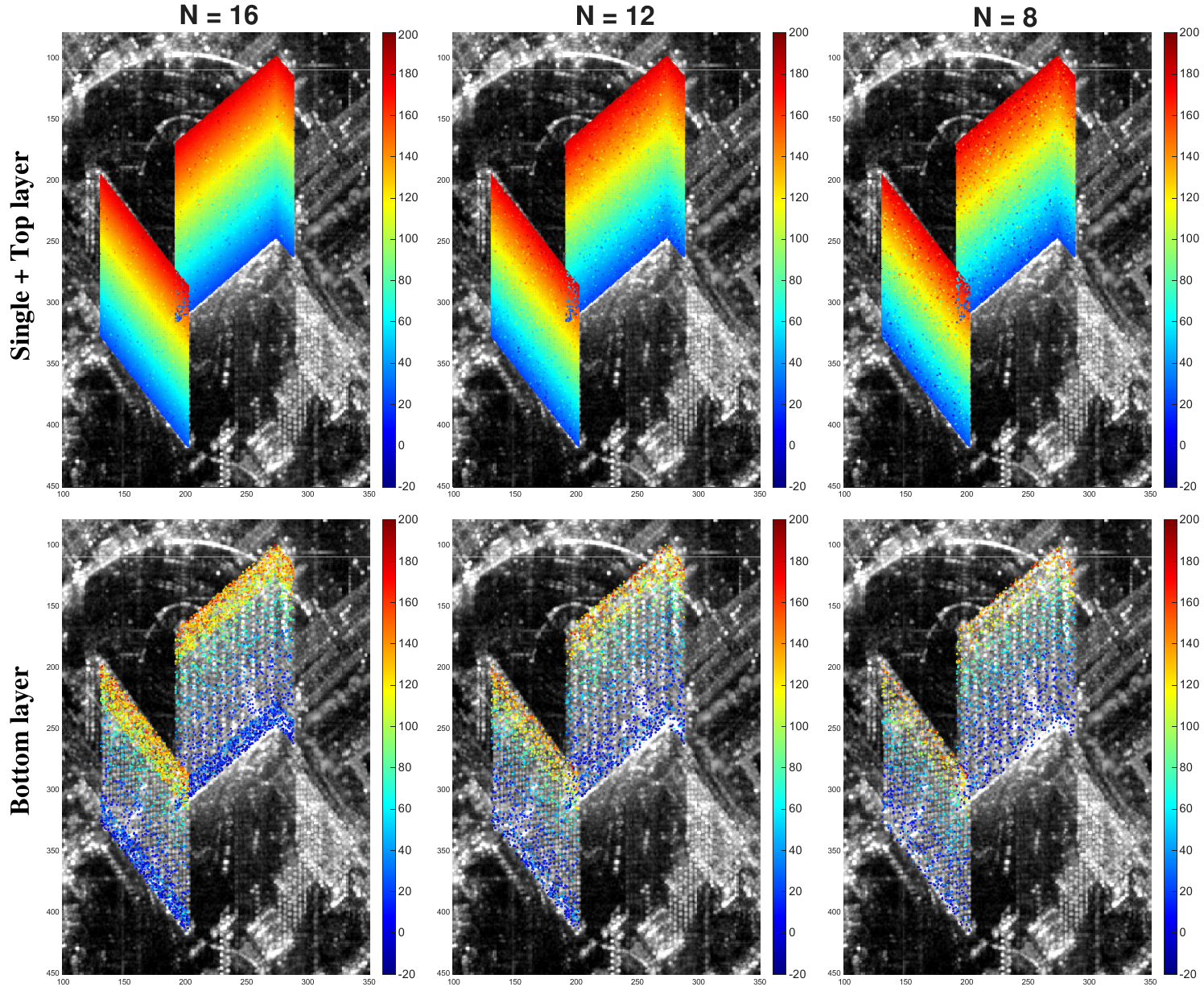}
\caption{Real-data reconstruction obtained by DUSG-Tomo-Net using different numbers of CH-1 acquisitions. From left to right, the columns correspond to $N=16$, $N=12$, and $N=8$. The top row combines the detected single-scatterer layer with the upper layer of the detected double-scatterer pixels, while the bottom row shows the corresponding lower layer.}
\label{fig:real_varyN}
\end{figure*}

To further examine adaptability to acquisition geometry using real data, we reconstruct the same scene using reduced acquisition subsets with $N=12$ and $N=8$, in addition to the full $N=16$ stack. For each reduced configuration, the two endpoint baselines are retained to preserve the physical aperture, while the remaining acquisitions are selected from the original baseline sequence. The same trained DUSG-Tomo-Net is used for all three configurations, with the geometry operator $\bPsi_{\mathcal G}$ recomputed according to the selected physical baselines. As shown in Fig.~\ref{fig:real_varyN}, the principal elevation structure remains consistent as the number of acquisitions is reduced from 16 to 12 and 8. The dominant facade geometry and its continuous elevation gradient are retained in all three configurations despite the reduced number of physical observations. The lower double-scatterer layer also preserves the main spatial pattern observed with the full stack, although the detections become progressively sparser as $N$ decreases, particularly for $N=8$. These results are consistent with the simulations in Section~IV-D: reducing the number of acquisitions decreases the available measurement information, but the same learned parameterization remains applicable after the geometry-dependent operator is updated.

\begin{figure*}[t]
\centering
\includegraphics[width=0.43\textwidth]{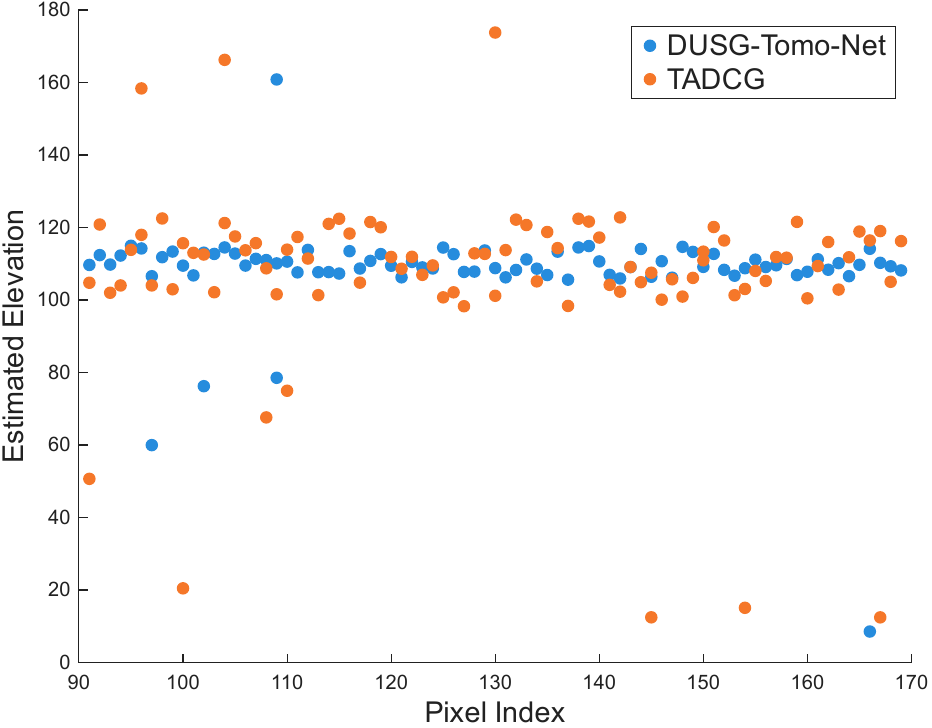}
\includegraphics[width=0.43\textwidth]{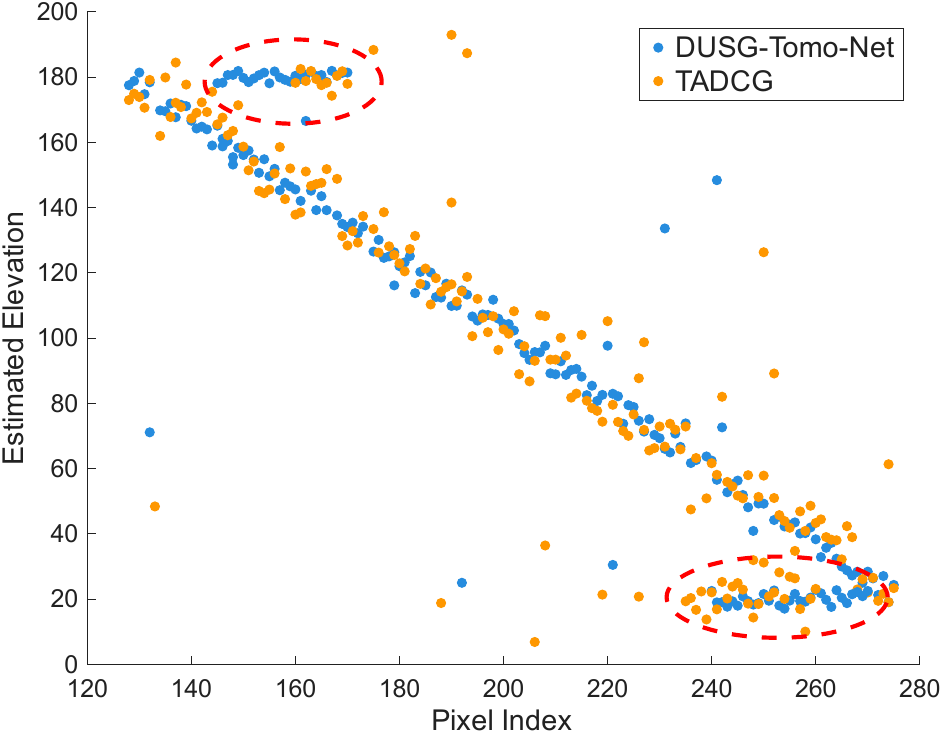}
\caption{Slice-wise elevation reconstruction of the real scene. Left: slice $F_1F_2$, approximately parallel to the lower edge of the building. Right: slice $R_1G_1$, approximately perpendicular to it. The vertical axis denotes the estimated elevation of the detected scatterers along each slice.}
\label{fig:real_slice_comparison}
\end{figure*}

To examine the reconstruction locally, two representative slices are extracted from the right-hand building: $F_1F_2$, approximately parallel to the lower edge of the building, and $R_1G_1$, approximately perpendicular to it. As shown in Fig.~\ref{fig:real_slice_comparison}, DUSG-Tomo-Net forms a tighter elevation band along $F_1F_2$, while TADCG exhibits greater dispersion and more isolated outliers. Along $R_1G_1$, both methods recover the dominant descending elevation trend, but DUSG-Tomo-Net follows the main branch more continuously. In the highlighted layover regions, its secondary-layer estimates extend over longer and more spatially coherent intervals, whereas the corresponding TADCG detections are more diffuse and intermittent.

\begin{figure}[t]
\centering
\includegraphics[width=0.98\columnwidth]{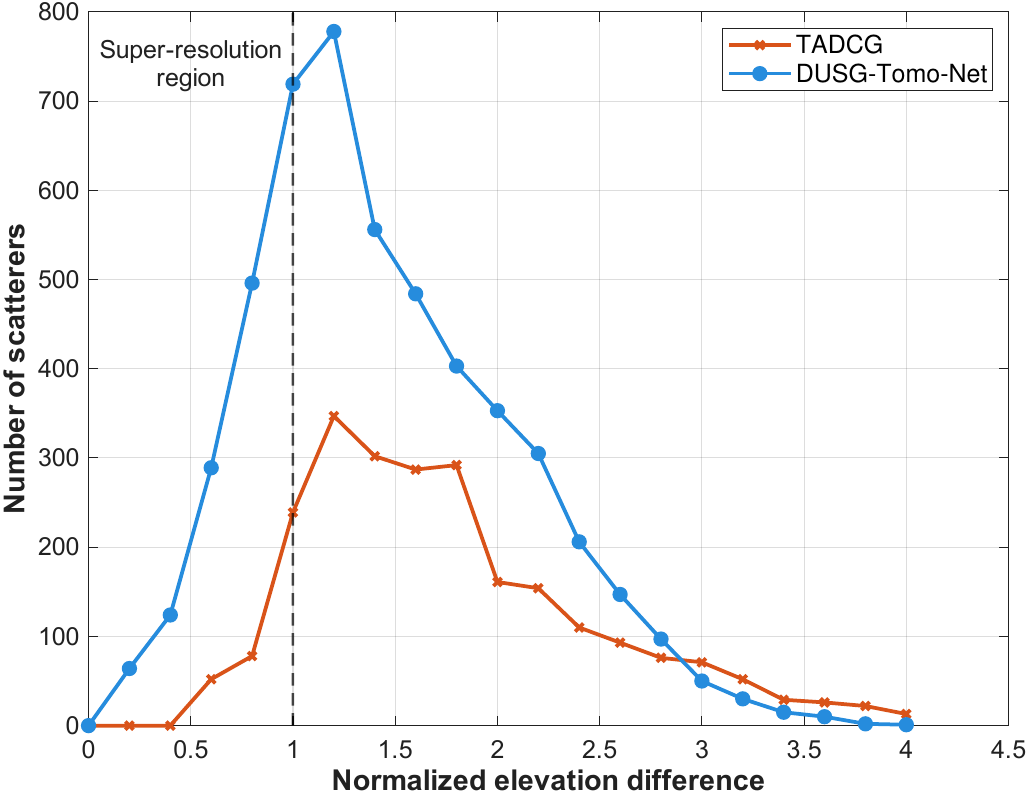}
\caption{Histogram of the normalized elevation separation between detected double-scatterer components. The dashed line denotes the Rayleigh boundary, and the region to its left corresponds to the sub-Rayleigh regime.}
\label{fig:real_ele_diff}
\end{figure}

Figure~\ref{fig:real_ele_diff} summarizes the separation statistics of the detected double-scatterer pixels using the full 16-image stack. DUSG-Tomo-Net identifies 5129 double-scatterer pixels, compared with 2404 for TADCG. Of these, 1692 DUSG-Tomo-Net detections and 369 TADCG detections lie at or below the Rayleigh boundary, corresponding to 33.0\% and 15.3\% of the respective double-scatterer detections. Thus, the larger number of double-scatterer responses obtained by DUSG-Tomo-Net is accompanied by a higher proportion of near- and sub-Rayleigh separations.

\section{Discussion}

\subsection{Single-Look Structured Estimation and Super-Resolution Behavior}

The proposed formulation operates on pairwise products obtained from a single complex TomoSAR observation vector. According to Eqs.~\eqref{eq:pairwise_observation}-\eqref{eq:lag_forward_model}, each physical-lag observation contains the desired lag-dependent component together with realization-dependent multi-scatterer cross terms,
signal-noise perturbations, and noise-noise perturbations. Under the random-phase model adopted in Section~III-A, these residual terms vanish in expectation for distinct acquisitions, yielding the ensemble covariance relation in Eq.~\eqref{eq:pairwise_expectation}. For an individual single-look realization, however, the residual perturbations remain and must be suppressed through structured estimation.

This observation also explains the SNR dependence of the super-resolution behavior in Section~IV-C. As the SNR decreases, the noise-related terms in the pairwise observations become more influential relative to the desired
lag-domain component, making the recovery of closely spaced spectral components less stable. DUSG-Tomo-Net addresses this problem by jointly estimating a common latent lag sequence from all admissible pairwise
observations and constraining the estimate through learned lag-domain regularization, analytical data consistency, and Toeplitz-PSD structure. Accordingly, increasing SNR leads to a progressively earlier transition to successful sub-Rayleigh separation and reduced elevation-estimation variance, as observed in Figs.~\ref{fig:superres_pd} and~\ref{fig:superres_accuracy}. However, the achievable super-resolution remains jointly determined by measurement quality and acquisition geometry. Low SNR, sparse baseline coverage, or unfavorable baseline sampling can reduce the information available for distinguishing closely spaced components.

\subsection{Geometry Decoupling, Virtual-Lag Design, and Applicability}

Geometry decoupling in DUSG-Tomo-Net refers to the architectural separation between the acquisition-dependent analytical measurement mapping and the geometry-shared learned regularization. The physical baseline configuration is explicitly represented by $\bPsi_{\mathcal G}$, whereas the learned modules operate on the fixed $N_v$-dimensional virtual-lag representation. Consequently, changing the baseline positions or the number of acquisitions does not require the dimensions or parameters of the learned regularizer to be modified.

The geometry-generalization experiments in Section~IV-D provide empirical evidence for this design. Under increasing perturbations of the reference nonuniform geometry, DUSG-Tomo-Net exhibits substantially less degradation than the geometry-coupled $\gamma$-Net, while remaining broadly comparable to geometry-matched model-based reconstruction. Similarly, when the number of acquisitions differs from the training configuration, the proposed method maintains performance comparable to SL1MMER without modifying its learned parameterization. The real-data experiment with reduced CH-1 acquisition subsets shows the same qualitative behavior: the dominant elevation structure is retained as the number of acquisitions decreases, although the secondary-layer reconstruction becomes progressively sparser as measurement information is reduced.

This reuse is subject to the representational compatibility conditions established in Section~III-C. For a prescribed virtual-lag representation $(N_v,\Delta_b)$, the physical baseline-difference range must satisfy
\begin{equation}
    B_{\max,\mathcal G}
    \leq
    (N_v-1)\Delta_b,
\end{equation}
while $\Delta_b$ must also satisfy the principal-branch requirement associated with the prescribed elevation support. These conditions determine whether a new acquisition geometry can be represented in the same virtual-lag domain, but they do not guarantee invariant reconstruction accuracy. Even for compatible geometries, performance can vary with physical-lag coverage, sampling density, and the conditioning of $\bPsi_{\mathcal G}$, as reflected by the geometry-to-geometry variations observed in Section~IV-D.

The parameters $N_v$ and $\Delta_b$ therefore involve a practical tradeoff. The virtual-lag span
\begin{equation}
    B_v=(N_v-1)\Delta_b
\end{equation}
should cover the largest baseline span expected for the intended deployment, whereas $\Delta_b$ must remain sufficiently small to avoid elevation ambiguity and to provide an adequate approximation of the continuous lag response. Increasing $N_v$ enlarges the representational range but also increases the cost of the Toeplitz-PSD projection, whose eigendecomposition scales cubically with $N_v$. Accordingly, a mission-oriented virtual-lag range with a moderate coverage margin is preferable to an unnecessarily large representation. The remaining architectural parameters, including the unfolding depth $M$, the number of Dykstra iterations $J$, and the compact CNN regularizer, are kept fixed across all experiments; they determine the tradeoff between reconstruction accuracy and computational cost rather than the physical geometry supported by the model.

\subsection{Practical Model-Order Handling and Evaluation Scope}

The reconstruction experiments in Sections~IV-C and IV-D are designed primarily to evaluate elevation localization and two-scatterer separation. The model order is therefore prescribed in the synthetic experiments so that reconstruction accuracy can be assessed independently of model-order estimation. Although the effective detection rate follows the criterion used in previous TomoSAR studies, it should be interpreted here as the proportion of trials for which the prescribed two-scatterer model is successfully reconstructed and localized within the specified tolerance. It is therefore different from a statistical detection probability evaluated at a controlled false-alarm probability, as used in GLRT-based TomoSAR detection.

For real data, the model order is unknown and varies spatially. In the present implementation, a practical screening-and-validation procedure is used for $K\in\{0,1,2\}$. Low-coherence pixels are first excluded from point-scatterer inversion. For the retained pixels, the eigenspectrum of the reconstructed Toeplitz-PSD matrix is used to identify plausible two-component responses. Root-MUSIC then estimates the elevations, which are subjected to physical validation. Candidate double-scatterer solutions that violate the admissible elevation range or minimum-separation requirement are rejected in favor of a single-scatterer interpretation.

\subsection{Computational Efficiency}

DUSG-Tomo-Net replaces convergence-driven optimization by a fixed number of unfolded layers, but it does not eliminate all cubic matrix operations. For each layer, the principal computational components are the compact one-dimensional CNN regularizer, the analytical data-consistency update, and the finite-step Toeplitz-PSD projection. With geometry-dependent factorizations precomputed for a fixed acquisition stack, the representative online complexity is
\begin{equation}
\mathcal{O}\!\left[
M\left(
k_c C^2 N_v
+
J N_v^3
+
N_v^2
\right)
\right],
\end{equation}
followed by the $\mathcal{O}(N_v^3)$ eigendecomposition required for the Root-MUSIC readout. The cubic term mainly arises from the PSD projection and therefore becomes increasingly relevant when a substantially larger virtual-lag dimension is adopted.

The principal practical advantage is the deterministic inference depth. Under the implementation used in this study, the measured per-pixel runtimes are approximately 2.3~ms for TADCG, 29.0~ms for SL1MMER, and 3.65~ms for DUSG-Tomo-Net. All runtime measurements were obtained in the same computational environment using an NVIDIA RTX 5090 GPU. Thus, TADCG remains the fastest implementation in absolute runtime, whereas DUSG-Tomo-Net is substantially faster than the iterative sparse-recovery implementation of SL1MMER and operates in the same millisecond-scale regime as TADCG. More importantly, its computational workload is fixed once $M$, $J$, and $N_v$ are specified and is amenable to parallel implementation across pixels.

\subsection{Limitations}

Several limitations define the current scope of DUSG-Tomo-Net. First, the model-order determination used in the real-data experiments remains empirical. The current screening-and-validation procedure combines coherence-based pixel selection, eigenspectrum information from the reconstructed Toeplitz-PSD matrix, and physical constraints on the Root-MUSIC solutions. Although this procedure provides a practical means of distinguishing the dominant single- and double-scatterer responses considered in this study, its decision thresholds are empirically specified and are not statistically calibrated. A more rigorous model-order selection strategy, for example by integrating the recovered structured representation with GLRT-based detection, information criteria, or other statistically controlled decision rules, would further improve the completeness of the reconstruction pipeline.

Second, the present formulation and experiments focus on sparse urban resolution cells dominated by a small number of coherent point-like scatterers, with at most two dominant components considered in the current implementation. The CH-1 experiments demonstrate the method primarily on building facades with pronounced layover and double-scattering responses. Distributed scattering, vegetation, and more complex mixtures containing a larger number of elevation components have not yet been systematically investigated. Extending the structured reconstruction and spectral readout to these more general scattering regimes remains an important direction for future work.

Finally, geometry decoupling should not be interpreted as unrestricted geometry or sensor invariance. The current experiments demonstrate model reuse across representationally compatible nonuniform baseline configurations and different acquisition numbers within the considered imaging setting. Reconstruction performance can still vary with physical-lag coverage, sampling density, and the conditioning of the geometry operator. Moreover, transfer across substantially different sensor modes, wavelength regimes, elevation supports, or baseline ranges beyond the prescribed virtual-lag domain has not been established and will require further investigation.

\section{Conclusion}

This paper presented DUSG-Tomo-Net, a geometry-decoupled deep unfolding framework for gridless super-resolution TomoSAR under nonuniform baselines. The proposed method reformulates single-look pairwise observations through a Toeplitz-compatible virtual-lag representation, explicitly incorporates the acquisition geometry through an analytical operator, and combines learned lag-domain regularization with geometry-dependent data consistency and Toeplitz-PSD structural enforcement. Continuous-valued scatterer elevations are subsequently recovered through Root-MUSIC without constructing a discretized elevation dictionary.

The experiments demonstrate that DUSG-Tomo-Net provides accurate continuous-domain localization and reliable near- and sub-Rayleigh double-scatterer reconstruction under nonuniform baselines. More importantly, the same learned parameterization remains applicable to the tested perturbed baseline configurations and acquisition numbers after the geometry-dependent operator is updated. Its performance is broadly comparable to that of geometry-matched SL1MMER and degrades substantially less than that of the geometry-coupled $\gamma$-Net. The CH-1 experiments further demonstrate spatially coherent urban elevation reconstruction and preservation of the dominant elevation structures when the acquisition stack is reduced from 16 to 12 and 8 images. Together with the deterministic fixed-depth inference, these results support geometry-decoupled virtual-lag unfolding as an effective framework for gridless TomoSAR reconstruction under varying nonuniform acquisition geometries.

The current implementation employs an empirically defined model-order screening procedure and focuses primarily on sparse urban scenes containing one or two dominant point-like components. Future work will therefore investigate statistically calibrated model-order selection, more general distributed and higher-order scattering regimes, and transfer across broader sensor and acquisition configurations.


%



\section*{Acknowledgment}
The authors are grateful to Prof. Gianfranco Fornaro of IREA-CNR, Italy, for his valuable suggestions, insightful comments, and stimulating discussions. They also thank Prof. Tao Li of Wuhan University, China, for providing the CH-1 SAR data used in the real-world experiments.

\ifCLASSOPTIONcaptionsoff
  \newpage
\fi



\bibliographystyle{IEEEtran}
\bibliography{paper.bib}
%



%

\begin{IEEEbiography}[{\includegraphics[width=1in,height=1.25in,clip,keepaspectratio]{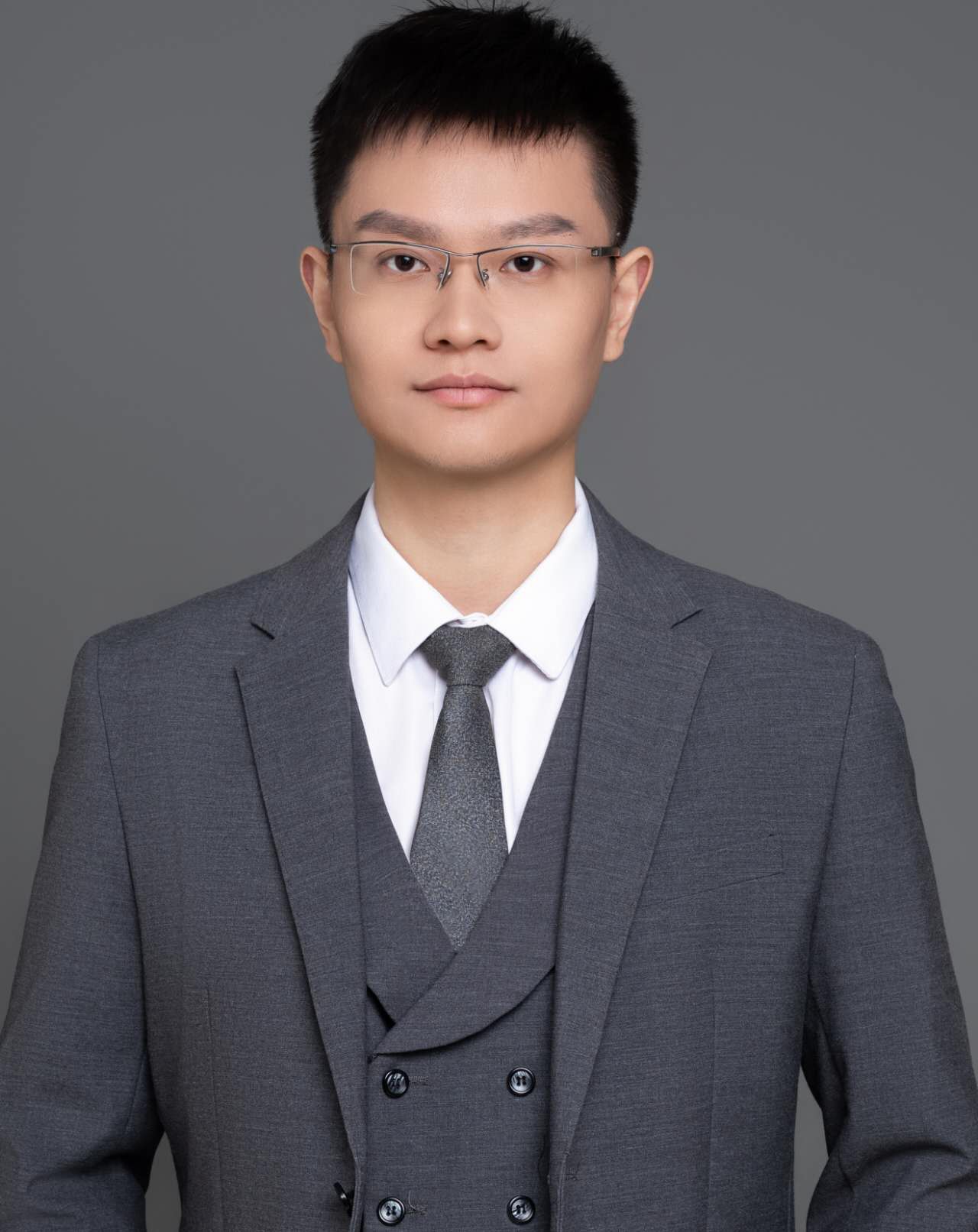}}]{Kun Qian}
received the Ph.D.\ degree in data science in Earth observation from the Technical University of Munich, Munich, Germany, in 2024. He received the B.Sc.\ degree in remote sensing and information engineering from Wuhan University, Wuhan, China, and the B.Sc.\ degree in aerospace engineering and geodesy from the University of Stuttgart, Stuttgart, Germany, both in 2017. He received the M.Sc.\ degree in aerospace engineering and geodesy from the University of Stuttgart in 2018. His research interests include data-driven methods, deep unfolding algorithms, and their applications to multibaseline SAR tomography and multitemporal InSAR.
\end{IEEEbiography}

\begin{IEEEbiography}[{\includegraphics[width=1in,height=1.25in,clip,keepaspectratio]{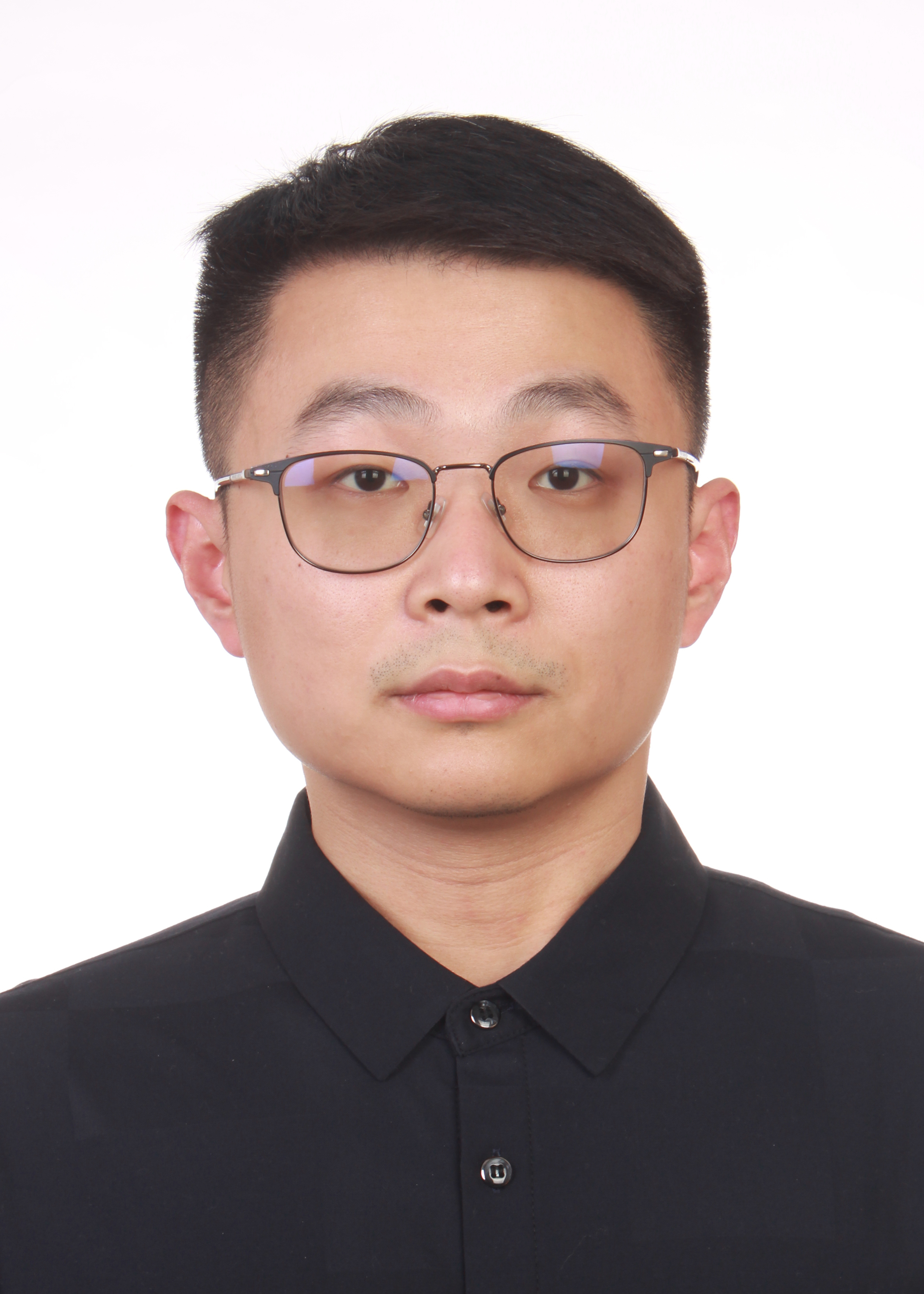}}]{Zhuge Xia}
received the Ph.D.\ degree in civil engineering and geodetic science from Leibniz University Hannover, Hannover, Germany, in 2023. From 2019 to 2023, he conducted his doctoral research at the GFZ German Research Centre for Geosciences, Potsdam, Germany. He received the B.Sc.\ degree in surveying and mapping engineering from Wuhan University, Wuhan, China, and the B.Sc.\ degree in aerospace engineering and geodesy from the University of Stuttgart, Stuttgart, Germany, both in 2017. He received the M.Sc.\ degree in aerospace engineering and geodesy from the University of Stuttgart in 2018. His research interests include satellite remote sensing, machine learning methods, and applications to geohazard monitoring and modeling.
\end{IEEEbiography}

\begin{IEEEbiography}[{\includegraphics[width=1in,height=1.25in,clip,keepaspectratio]{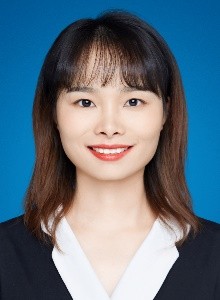}}]{Qian Ma}
received the Ph.D.\ degree from Sun Yat-sen University, Shenzhen, China, in 2025 and the M.S.\ degree from the National University of Defense Technology, Changsha, China, in 2021. She is currently a lecturer with the School of Medical Information Engineering, Guangzhou University of Chinese Medicine. Her research interests include 3-D imaging, inverse problems, and deep learning, with a particular focus on TomoSAR-based 3-D imaging of urban buildings.
\end{IEEEbiography}

\begin{IEEEbiography}[{\includegraphics[width=1in,height=1.25in,clip,keepaspectratio]{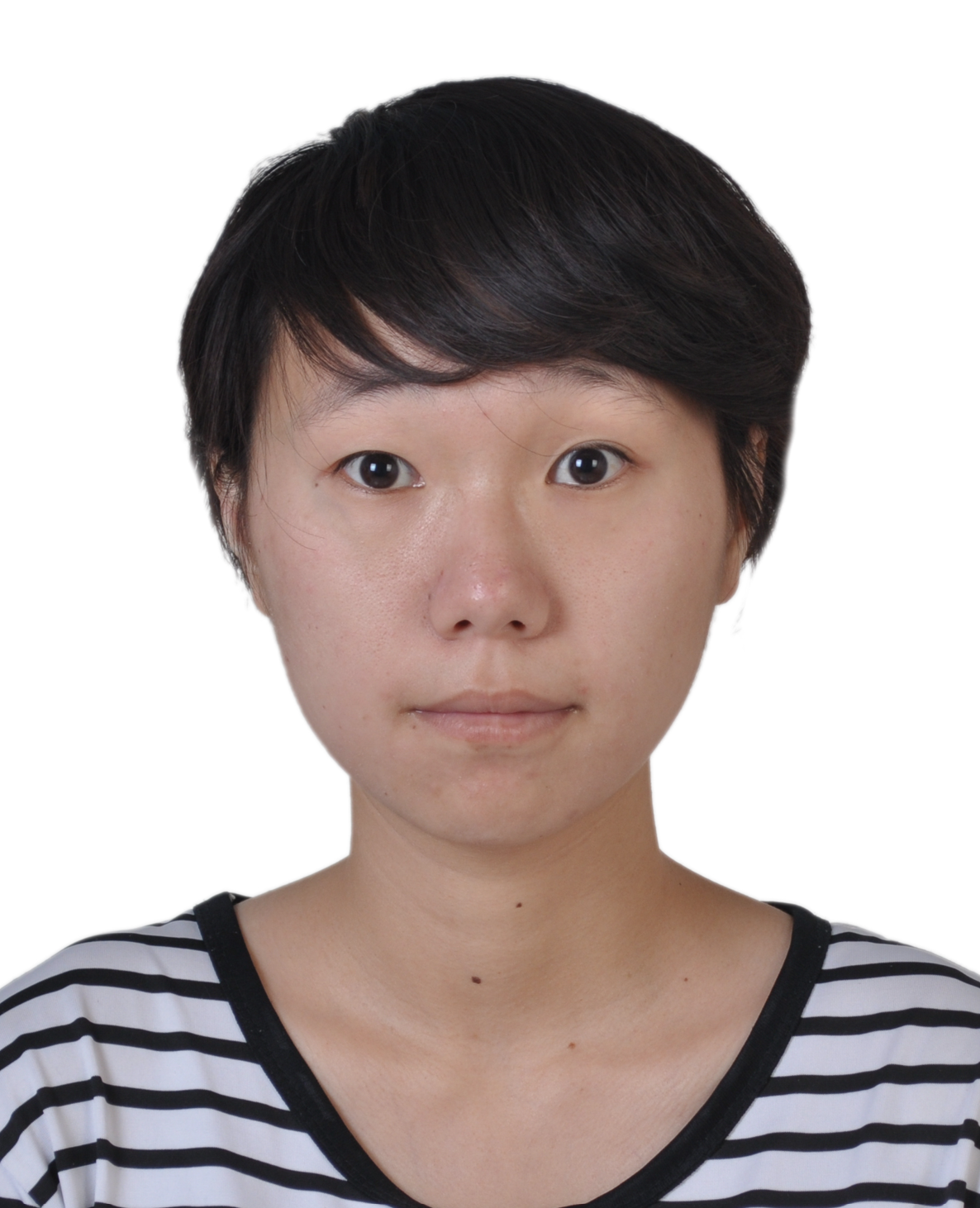}}]{Qi Zhang}
received the Ph.D.\ degree in geospatial engineering from the University of New South Wales (UNSW), Sydney, Australia, in 2022. She received the B.E.\ and M.E.\ degrees in information engineering from the Beijing Institute of Technology (BIT), Beijing, China, in 2014 and 2016, respectively. She is currently a postdoctoral researcher at the Technical University of Munich, Munich, Germany. Her research interests include deep-learning-based algorithm design and its applications to Earth observation using SAR and multispectral remote sensing.
\end{IEEEbiography}

\begin{IEEEbiography}[{\includegraphics[width=1in,height=1.25in,clip,keepaspectratio]{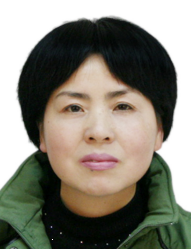}}]{Xiufeng He (Senior Member, IEEE)} received the B.Eng.\ and M.S.\ degrees in control and navigation from Nanjing University of Aeronautics and Astronautics, Nanjing, China, in 1986 and 1989, respectively, and the Ph.D.\ degree in navigation and survey engineering from The Hong Kong Polytechnic University, Hong Kong, in 1998. From 1998 to 1999, she was a postdoctoral fellow at the Norwegian University of Science and Technology, Trondheim, Norway, where she worked on guidance, navigation, and control. In 2000, she was a research fellow with the Global Positioning System (GPS) Center, Nanyang Technological University, Singapore. She is currently a professor and the director of the Institute of Satellite Navigation and Spatial Information, Hohai University, Nanjing, China. She has authored or coauthored three books and more than 200 refereed papers. Her research interests include interferometric synthetic aperture radar, global navigation satellite systems (GNSS), inertial navigation systems (INS), and low-cost integrated GNSS/INS navigation systems.
\end{IEEEbiography}







\end{document}